\documentclass[a4paper,12pt]{article}
\usepackage[T1]{fontenc}
\pdfoutput=1

\usepackage{jheppub}
\usepackage{macro}
\usepackage{appendix}
\usepackage{comment}
\usepackage{array}
\usepackage{yfonts}

\title{\textbf Axialgravisolitons at infinite corners}

\author[a,b]{Federico Manzoni}

\affiliation[a]{Mathematics and Physics department, Roma Tre, Via della Vasca Navale 84, Rome, Italy}
\affiliation[b]{INFN Roma Tre Section, Physics department, Via della Vasca Navale 84, Rome, Italy}
\emailAdd{federico.manzoni@uniroma3.it, ORCID ID: 0000-0002-9979-6154}

\abstract{Gravitational solitons (gravisolitons) are particular exact solutions of Einstein field equation in vacuum build on a given background solution. Their interpretation is not yet fully clear but they contain many of the physically relevant solutions low $N$-solitons solutions.  
However, a systematic study and characterization of gravisolitons solution for every $N$ is lacking and their relevance in a theory of quantum gravity is not fully understood. This work aims to investigate and characterize some properties of $N$-axialsoliton solutions such as their asymptotically behaviour and asymptotic symmetries given minimal assumptions on the background metric. We develop an explicit systematic asymptotically expansion for the $N$-axialsoliton solution and we compute the leading order of the asymptotic killing vectors. Moreover, in the perspective to better understand the role of gravisolitons in quantum gravity we make a link, and a one of the first explicit test, to the corner symmetry proposal deriving which subalgebra of the universal corner symmetry algebra is generated by the asymptotic Killing vectors of $N$-axialsoliton solution. In the spirit of the corner proposal, the axialgravisoliton corner symmetry algebra (\textfrak{agcsa}) can be useful for the quantization of the non-asymptotically flat sector of gravity while, in the spirit of IR triangle, new soft theorems and memory effects could emerge. 
\\
\\
\\
\textit{Published in Classical and Quantum Gravity 41 177001.}}

\begin{document}

\maketitle
%\flushbottom
%\newpage

\section{Introduction}
Solitons were first observed in 1834 by the Scottish engineer John Scott Russell in the Union Canal. After the discovery of soliton solutions in hydrodynamic equations, such as KdV equation, the study of these solutions led to the development of new powerful techniques with the aim of having systematic procedures for the construction of soliton solutions. One of these techniques is the spectral transform, or in a better way, the Inverse Scattering Method (ISM) that was introduced in 1967 by Gardner, Greene, Miura and Kruskal \cite{grenne} to solve the problem at the initial values of the KdV and then extended to other situations of interest, for example as did by Zakharov and Shabat in 1972 \cite{shabat} with the non-linear Schrödinger equation. The extension of the spectral transform in the realm of Einstein gravity happened in 1978 by Belinskii and Zakharov \cite{1978ZhETF..75.1953B} and two years later Alekseev proposed the extension to the Maxwell-Einstein theory \cite{1980JETPL..32..277A}.

Spectral transform method is essentially based on the possibility of writing the non-linear equation under consideration as a condition of integrability of an associated matrix system of linear differential equations, the so-called the Lax pair of the problem. Once the appropriate Lax pair has been determined, the problem is divided into two parts: the direct problem and the inverse problem. Let us highlights the most important steps:
\begin{enumerate}
    \item  find the Lax pair, that is two linear operators $L$ and $M$ such that $Lv=\lambda v$ and $\partial_t v=Mv$. It is extremely important that the eigenvalue $\lambda$ be independent of time (isospectrality). Necessary and sufficient conditions for this to occur is given by the so-called Lax equation, $\partial_tL+LM-ML=0$. After finding the appropriate Lax pair it should be the case that Lax equation recovers the original non-linear PDE;
    \item determine the time evolution of the eigenfunctions associated to each eigenvalue $\lambda$, the normalization constants and the reflection coefficient; these form the so-called scattering data. This time evolution is given by a system of linear ordinary differential equations. This step is the direct problem.
    \item  solving the Gelfand–Levitan–Marchenko (GLM) integral equation, a linear integral equation, to obtain the final solution of the original non-linear PDE. All the scattering data is required in order to do this. If the reflection coefficient is zero, the process becomes much easier. This step is the inverse problem.
\end{enumerate}
The procedure is schematized as
\begin{equation}
\begin{aligned}
& \ \ \ \ \ \ \ \ \ \ \ \ \ \ \ \ \ \ \ \ \  \ \ \ \ u(z,0) \xrightarrow{direct \ problem} S(\lambda,0)\\
&non linear \ evolution \Bigg\downarrow\ \ \ \ \ \ \ \ \ \ \ \ \ \ \ \ \ \ \ \ \ \ \ \   \Bigg\downarrow \ linear \ evolution \\
&\ \ \ \ \ \ \ \ \ \ \ \ \ \ \ \ \ \ \ \ \  \ \ \ \ u(z,t) \xleftarrow{inverse \ problem } S(\lambda,t)
\label{TS}
\end{aligned}    
\end{equation}
where $u(z,t)$ is the quantity to be determined, $u(z,0)$ its initial condition, $S(\lambda,0)$ are the scattering data at $t=0$ and $S(\lambda,t)$ are the evolved scattering data.
As can be seen from schematization \ref{TS}, the spectral transform method can be thought as a non-linear analogue, and in some sense generalization, of the Fourier transform. 

Soliton solutions emerge when the reflection coefficient vanishes, in that case the GLM integral equation reduces to an algebraic system. The discrete elements $\lambda_k$ of the spectrum are intimately related to solitons, indeed given a background solution, i.e. the simplest one, of our problem we can define a new solution 
\begin{equation}
    \psi=\chi\psi_b
\end{equation}
where $\psi_b$ is called generating matrix and
\begin{equation}
    \chi=\mathcal{I}+\sum_{k}\frac{O_k}{\lambda-\lambda_k}
\end{equation}
is called the dressing matrix; $O_k$ are operators do not depend on the spectral parameter. Note that the dressing matrix has pole exactly on the value of the discrete eigenvalues and a $N$ pole dressing matrix can generate a $N$-soliton solution. 

In this work we focus on gravitational solitons, also known as gravisolitons. These are particular exact solutions of the Einstein, or Einstein-Maxwell, field equations that can be constructed by the use of the inverse scattering method known, respectively as Belinski–Zakharov (BZ) transform \cite{1978ZhETF..75.1953B} and  Belinski-Zakharov-Alekseev (BZA) transform \cite{1980JETPL..32..277A}. Their interpretation is not fully clear but black holes and cosmological solutions are special cases of gravitational solitons; moreover, recently, also more exotic solutions are determined using generating methods like the one presented above \cite{Vigano:2022hrg}. Despite the term 'soliton' being used to describe gravisolitons, their behavior is very different from other solitons; in particular, gravisolitons do not preserve their amplitude and shape in time and in scattering processes. Recently 
\cite{Asselmeyer_Maluga_2020} has been proposed, with some consistency tests, that gravisolitons have many properties of dark matter, such as no interaction with electromagnetic field but act on matter via gravitation. Moreover the authors showed that the gravitational lensing effect of gravisolitons agreed with the lensing effect of usual matter and that they have the same equation of state $w = 0$ as matter has. Moreover, in \cite{mcnamara2021gravitational}, the author shows how gravitational solitons that can be considered as topological, in the sense of some non-trivial homotopy groups, naturally carry charges beyond those of any local excitation of the quantum fields. Therefore, there are symmetries of the EFT which are broken as soon as the topology of space-time is allowed to fluctuate, even without any additional UV degrees of freedom. The author finds that the effect of gravitational solitons is to break the non-invertible symmetry to the maximal group-like subsymmetry and if we further demand that remaining group-like symmetry is broken by additional degrees of freedom, we find a complete spectrum. Hence, the completeness hypothesis follows from the absence of group-like symmetries once gravitational solitons are taken into account. Therefore gravisolitons can play a fundamental role in a theory of quantum gravity but this role is mysterious at the moment and a more deep understanding of gravisolitons could shed new light on the issue. 

The most recent developments in the theory of gravisolitons are mainly contained in \cite{kordas2011transition, durgut2022supersymmetric, Durgut_2023, Durgut_20232}; in the first the hamiltonian methods are applied to gravisolitons and it is proved that the transition matrix satisfies equations familiar from integrable PDEs while the others deal with the study of gravisolitons in Einstein theory of negative cosmological constant constructing  supersymmetric gravitational soliton solutions of five-dimensional gauged supergravity coupled to arbitrarily many vector multiplets and studying the behaviour of causal geodesics and thermodynamic properties of Eguchi-Hanson-AdS$_5$ gravisolitons.
Therefore, motivated by the recent partially lacking, a part \cite{kordas2011transition, durgut2022supersymmetric, Durgut_2023, Durgut_20232}, of formal and mathematical characterizations of gravisoliton solutions and by the possible central role of them in a theory of quantum gravity, we are interested in study some of their properties, from a formal and mathematical point of view, focusing of gravisolitons with axial symmetry or axialgravisolitons. We first search and develop a systematic expansion for the $N$-axialgravisoliton metrics to better understand some of their asymptotic properties. This expansion can play, for $N$-axialgravisoliton space-time, a similar role of Bondi expansion for asymptotically flat space-time and Fefferman-Graham expansion for asymptotically AdS space-time. As the $N$-axialgravisoliton solution strongly depends on the background solution also its asymptotic expansion depends on it and, more specifically, depends on its radial expansion which is divided into cases as we will see in Chapter \ref{par3}. As first application of the aforementioned expansion, we investigate the asymptotic symmetries of the $N$-axialgravisoliton solution. The first example of asymptotic symmetry in gravity is due to the pioneering work of Bondi, Metzner, Van der Burg and Sachs \cite{Bondi:1962px, Sachs:1962wk, PhysRev.128.2851, Compere:2018aar} which found the group of asymptotic physical symmetries, even on empty Minkowski space-
time, is not Poincaré but rather the infinite-dimensional Bondi–van der Burg–Metzner–Sachs (BMS) group. Recently it was shown how supertranslations at null infinity can be achieved in the convolutional double copy framework \cite{Ferrero:2024eva}. However, asymptotic symmetries find place not only in gravity but also in other gauge theories \cite{Strominger:2017aeh, Francia:2018jtb, Campoleoni:2018uib, Pasterski:2015zua},  like standard Maxwell and Yang-Mills theories and more exotic gauge theories like $p$-forms and mixed symmetry tensors \cite{Henneaux:1986ht, Afshar:2018apx, Esmaeili:2020eua, manz1, manz2}. Thanks to the asymptotic expansion we developed, we compute, in Paragraph \ref{subpar3.2}, the leading order Killing vectors and study their bracket and therefore their algebra. In the same Paragraph we make a link, and one of the first explicit test, to the corner proposal which is a novel way to deal with quantum gravity \cite{Speranza:2017gxd, Geiller:2017whh, Donnelly:2020xgu, Ciambelli:2021vnn, Ciambelli:2021nmv, Canepa:2022uii, ciambelli2023asymptotic} where corners, i.e. codimension two embedded manifold and the symmetry algebra at corners, are lifted to be fundamental ingredients of any theory of quantum gravity. We find that for every $N$, the axialgravisoliton corner symmetry algebra (\textfrak{agcsa}) is a subalgebra of the proposed universal corner symmetry algebra. This suggests a positive feedback to testing the corner proposal explicitly and open the way to a possible quantization to the non-asymptotically flat and non-perturbative sector of gravity by studying the representations of \textfrak{agcsa}.

\section{Axially symmetric gravitational solitons}
Let us discuss axially symmetric gravitational solitons \cite{Belinski:2001ph},\cite{Manzoni:2021dij}. First of all, we need to taken into account an axisymmetric stationary background metric, which therefore admits two Killing fields: one is space-like $m=\partial_{\phi}$ and the other is time-like $k=\partial_{t}$.  We will adopt the notation $(x^0,x^1,x^2,x^3)=(t,\phi,\rho,z)$ for the coordinates and comma notation for derivatives.\\
Since we are at our disposal two Killing fields we can eliminate the dependence on one spatial and one temporal variables, hence metric components depend only on $\rho$ and $z$. However, we can simplify further invoking diffeomorphisms invariance of general relativity. In four dimensional space-time this symmetry allows us to impose four conditions on the metric without loss of generality: we will impose that
\begin{equation}
    (g_b)_{22}=(g_b)_{33}, \ (g_b)_{02}=0, \ (g_b)_{12}=0, \ (g_b)_{23}=0.
\end{equation}
However, to use spectral transform method we need one more simplification and since we exhausted our freedom
to manipulate the metric without loss of generality we must impose a physical
conditions $(g_b)_{03}=(g_b)_{13}=0$; in the and we can write the metric in the form
\begin{equation}
    ds^2=(g_b)_{22}(d\rho^2+dz^2)+(g_b)_{ij}dx^idx^j,
\label{MVAS}    
\end{equation}
where, $i,j=0,1$ and where $g_{22}$ and $g_{ij}$ depend only on $x^2=\rho$ and $x^3=z$. For stationary metrics of the type \ref{MVAS} it can be imposed, without loss of generality, that the $2 \times 2$ block $(g_b)_{ij}$ has $det((g_b)_{ij})=-\rho^2$. 

Einstein field equations in vacuum for metric \ref{MVAS} can be rewritten as
\begin{equation}
\begin{aligned}
    &(\rho g_{,\rho}g^{-1})_{,\rho}+(\rho g_{,z}g^{-1})_{,z}=0,\\
    &(ln(g_{22}))_{,\rho}=-\frac{1}{\rho}+\frac{1}{4\rho}(U^2-V^2) \ , \ (ln(g_{22}))_{,z}=\frac{1}{2\rho}UV;
\label{EVAS}    
\end{aligned}    
\end{equation}
where 
\begin{equation*}
    U:=\rho g_{,\rho} g^{- 1}, \ V:=\rho g_{,z}g^{-1}
\end{equation*}
The $N$ axisymmetric solitons solution is
\begin{equation}
    ds^2=g_{22}^{(f)}(d\rho^2+dz^2)+g_{ij}^{(f)}dx^idx^j
\label{VB}    
\end{equation}
with
\begin{equation}
\begin{aligned}
    &g_{22}^{(ph)}=16C(g_{b})_{22}\rho^{-\frac{N^2}{2}}\bigg(\prod_{k=1}^N\lambda_k\bigg)^{N+1}\bigg(\prod_{k>l=1}^N\frac{1}{(\lambda_k-\lambda_l)^2}\bigg)det(\Gamma);\\
    &g_{ij}^{(ph)}=\pm \frac{(g_b)_{ij}}{\rho^N}\bigg(\prod_{k=1}^N \lambda_k\bigg)=\pm \frac{\big((g_b)_{ij}-\sum_{k,l=1}^ND_{kl}\lambda_k^{-1}\lambda_l^{-1}L_i^{(k)}L_j^{(l)}\big)}{\rho^N}\bigg(\prod_{k=1}^N \lambda_k\bigg),
\label{corfs}    
\end{aligned}    
\end{equation}
where 
\begin{equation}
\begin{aligned}
&D_{kl}:=(\Gamma^{-1})_{kl}; \ \ \ \ \Gamma_{kl}:=m^{(k)}_i(g_b)_{ij}m_j^{(l)}(\rho^2+\lambda_k \lambda_l)^{-1};\\
&L_i^{(k)}:=m_j
^{(k)}(g_b)_{ji}; \ \ \ \ m_i^{(k)}:=m_{2j}^{(k)}(\psi_b^{-1}(\lambda_k,\rho,z))_{ji}
\end{aligned}    
\end{equation}
where $m_{2j}^{(k)}$ are arbitrary constants and $\psi^{-1}(\lambda, \rho, z)$ is the inverse of the generating matrix solution of the PDE coupled system\footnote {To obtain the background generating matrix just replace the matrices $\hat{U}$ and $\hat{V}$ with their background counterpart, that is $U_b:=\rho g_{b,\rho}g_b^{-1}$ and $V_b:=\rho g_{b,z}g_b^{-1}$.}
\begin{equation}
\begin{aligned}
\bigg(\partial_z-\frac{2\lambda^2}{\lambda^2+\rho^2}\partial_{\lambda}\bigg)\psi=\frac{\rho V-\lambda U}{\lambda^2+\rho^2}\psi \ , \ \bigg(\partial_{\rho}+\frac{2\lambda \rho}{\lambda^2+\rho^2}\partial_{\lambda}\bigg)\psi=\frac{\rho V+\lambda U}{\lambda^2+\rho^2}\psi.   
\label{BNM}
\end{aligned}    
\end{equation}
The generating
matrix satisfy the relation 
\begin{equation}
    g(\rho,z)={\psi}(0,\rho,z);
\label{D}    
\end{equation}
therefore
\begin{equation}
 {\psi}(\lambda,\rho,z)=g(\rho,z)+\sum_n \lambda^n F(\rho,z).
 \label{D2}
\end{equation}
Furthermore, $C$ is an arbitrary constant but with a sign such that we have $g_{22}^{(ph)} \geq 0$; the $+$ or $-$ sign in front of the second of the \ref{corfs} must be chosen appropriately in order to have the correct signature of the metric. The lorentzian manifold with metric components given by \ref{corfs} is a physically space-time which is an exact, hence non-perturbative, solution of Einstein field equation in vacuum without cosmological constant. Solitons of the form \ref{VB} are called stationary axial symmetric solitons or axialgravisolitons.

Equations that determine the trajectories of the poles $\lambda_k$ are given by 
\begin{equation}
\begin{aligned}
&\lambda_{k,z}=\frac{-2\lambda_k^2}{\lambda_k^2+\rho^2},\\
&\lambda_{k,\rho}=\frac{2\lambda_k\rho}{\lambda_k^2+\rho^2};
\label{zzz}
\end{aligned}
\end{equation}
 and the solution of these equations is given by the solutions of the quadratic equation
\begin{equation}
    \lambda_k^2+2z\lambda_k-\rho^2=2w_k\lambda_k,
\label{EQ2}    
\end{equation}
where the $w_k$ are complex arbitrary constants. Poles are than given by
\begin{equation}
    \lambda_k=(w_k-z)\pm \sqrt{(w_k-z)^2+\rho^2},
\label{zzzz}    
\end{equation}
as one can see the root is always positive for real poles ($w_k$ real) and therefore the solitonic solution will not experience discontinuity and it will be present in all space-time in contrast to the case of not stationary gravisolitons.

The condition $det(g)=-\rho^2$, necessary to have a physical solution, be satisfied. To obtain the physical solution, $g^{(f)}$, we need to compute the determinant of the solution $g$; the calculation leads to
\begin{equation}
    det(g)=(-1)^N\rho^{2N}\bigg(\prod_{k=1}^N\frac{1}{\lambda_k^2}\bigg)det(g_b).
\label{der}    
\end{equation}
The \ref{der} shows that if the background solution is such that $det(g_b)=-\rho^2$ then the solution must necessarily be an even number of solitons, $N=2n$, otherwise the sign of the determinant of $g$ would change and this would lead to a non-physical metric. Hence, the simplest solution is the one with $N=2$ solitons and can be shown \cite{Manzoni:2021dij} that tuning in a suitable way parameters this solution returns Kerr black hole and more generally a Taub-NUT space-time \cite{Belinski:2001ph}.

\section{Asymptotic behavior and asymptotic symmetries}\label{par3}
Let us consider the axisymmetric gravisolitons metric \ref{corfs}, explicitly
\begin{equation}
\begin{aligned}
 &g_{22}^{(ph)}=16C(g_{b})_{22}\rho^{-\frac{N^2}{2}}\bigg(\prod_{k=1}^N\lambda_k\bigg)^{N+1}\bigg(\prod_{k>l=1}^N\frac{1}{(\lambda_k-\lambda_l)^2}\bigg)\det(\Gamma);\\
    &g_{00}^{(ph)}=\pm \frac{(g)_{00}}{\rho^N}\bigg(\prod_{k=1}^N \lambda_k\bigg)=\pm \frac{\big((g_b)_{00}-\sum_{k,l=1}^ND_{kl}\lambda_k^{-1}\lambda_l^{-1}L_0^{(k)}L_0^{(l)}\big)}{\rho^N}\bigg(\prod_{k=1}^N \lambda_k\bigg); \\
    &g_{11}^{(ph)}=\pm \frac{(g)_{11}}{\rho^N}\bigg(\prod_{k=1}^N \lambda_k\bigg)=\pm \frac{\big((g_b)_{11}-\sum_{k,l=1}^ND_{kl}\lambda_k^{-1}\lambda_l^{-1}L_1^{(k)}L_1^{(l)}\big)}{\rho^N}\bigg(\prod_{k=1}^N \lambda_k\bigg); \\
    &g_{10}^{(ph)}=g_{01}^{(ph)}=\pm \frac{(g)_{10}}{\rho^N}\bigg(\prod_{k=1}^N \lambda_k\bigg)=\pm \frac{\big((g_b)_{10}-\sum_{k,l=1}^ND_{kl}\lambda_k^{-1}\lambda_l^{-1}L_1^{(k)}L_0^{(l)}\big)}{\rho^N}\bigg(\prod_{k=1}^N \lambda_k\bigg),
\label{metrica}    
\end{aligned}    
\end{equation}
Our task is to develop an asymptotic expansion of this metric in order to study the subgroup of diffeomorphisms which take the asymptotic metric invariant. In cylindrical coordinates the asymptotic expansion is performed taking the limits
\begin{equation}
    \rho \rightarrow \infty, \ |z| \rightarrow \infty \ \  \mathrm{s.t.} \ \ \frac{\rho}{z} \sim O(1);
\end{equation}
however, for some reasons it is simpler taking limits in spherical coordinates\footnote{The change of coordinates is given by $(\phi,\rho,z) \rightarrow (\phi=\phi,r=\sqrt{\rho^2+z^2}, \theta=acrtang\big(\frac{\rho}{z}\big))$, or in other terms $\rho=rsin(\theta), \phi=\phi, z=rcos(\theta)$.}
\begin{equation}
    r\rightarrow \infty.
\end{equation}
It is therefore necessary to study the asymptotic behavior of the various factors appearing in \ref{metrica}. \\

We will not fix the background metric in order to work in full generality
\begin{equation}
    ds^2=(g_b)_{22}(d\rho^2+dz^2)+(g_b)_{ij}dx^idx^j \ \ \ i,j=0,1;
    \label{metric}
\end{equation}
we only adopt the necessary requests for consistency on $(g_b)_{ij}$: a symmetric matrix with determinant given by $det(g_b)=-\rho^2$
\begin{equation}
    (g_b)_{ij}=\begin{bmatrix}
(g_b)_{00}:=A & (g_b)_{01}:=B \\
(g_b)_{10}:=B & (g_b)_{11}:=C  
\end{bmatrix} \ \ \mathrm{s.t.} \ \ det((g_{b})_{ij})=AC-B^2=-\rho^2.
\end{equation}
Since we are interested in asymptotic behaviours, let us give an asymptotic expansion for elements $A,B,C$. We have assumed that background metric has a time-like Killing vector hence the metric is stationary; this implies we can define Komar-like quantities of the background metric. Despite these quantities could not have physical reasoning they are, however, conserved quantities like energy and angular momentum associate to the two Killing vectors of the background metric; therefore, under the physical requirement that conserved quantities do not diverge at infinity we can restrict the possible choice of elements $A,B,C$. 

Let us now switch to spherical coordinates $(t,\phi,r,\theta) \equiv (0,1,2,3)$; the background metric reads
\begin{equation}
    ds^2=(g_b)_{22}(dr^2+r^2d\theta^2)+(g_b)_{ij}dx^idx^j \ \ \ i,j=0,1;
    \label{metrica2}
\end{equation}
and metric elements are now function of $r$ and $\theta$.

Komar-like quantities are proportional to the surface integral of Killing forms, hence
\begin{equation}
\begin{aligned}
K_t \propto \int_{\partial S}*dk, \ \ \ K_{\phi} \propto \int_{\partial S}*dm;
\end{aligned}    
\end{equation}
as shown in Appendix \ref{appA} some combination of the background metric elements have to be a specific asymptotic behavior in order to make finite Komar-like quantities and therefore in order to make finite conserved quantities:
\begin{equation}
\begin{aligned}
&\frac{C}{ sin^2(\theta)}\sqrt{\frac{C}{(g_b)_{rr}}}\frac{\partial A}{\partial r}\sim r^{n_1}f_1(\theta);\\
& \frac{B}{ sin^2(\theta)}\sqrt{\frac{C}{(g_b)_{rr}}}\frac{\partial C}{\partial r} \sim r^{n_2}f_2(\theta);\\
&\frac{C}{ sin^2(\theta)}\sqrt{\frac{C}{(g_b)_{rr}}}\frac{\partial B}{\partial r} \sim r^{n_3}f_3(\theta);
\label{3.8}
\end{aligned}    
\end{equation}
with 
\begin{equation}
\begin{aligned}
n_1,n_2,n_3 \leq 1 
\end{aligned}
\end{equation}
and $f_1(\theta),f_2(\theta),f_3(\theta)$ arbitrary function of azimuthal angle that could be expanded in Legendre polinomials.

We now assume a power law\footnote{Recently also logarithmic terms are taken in account in these kind of expansions; would be interesting to study also this possibility.} behavior for the background metric elements when $r\rightarrow \infty$
\begin{equation}
\begin{aligned}
A \sim r^{\alpha}\mathcal{A}(\theta), \ \ B \sim r^{\beta}\mathcal{B}(\theta), \ \ C \sim r^{\gamma}\mathcal{C}(\theta), \ \ (g_b)_{rr} \sim r^{\delta}\mathcal{D}(\theta);
\end{aligned}
\end{equation}
the condition on the determinant can be made explicit in two ways as summarized in the following Table \ref{tab1}.
\begin{table}[ht]
\centering
\begin{tabular}{|l|l|}
\hline
  1 & $\alpha+\gamma=2, \ \ \ \beta=1, \ \ \ \mathcal{AC}(\theta)-\mathcal{B}^2(\theta)=-sin^2(\theta)$    \\
  \hline
  2    & $\alpha+\gamma=2, \ \ \ \mathcal{AC}(\theta)=-sin^2(\theta), \ \ \ \mathcal{B}^2(\theta)=0$   \\
 \hline 
 \end{tabular}
 \caption{\textit{Possible explicit choices for the determinant condition.}}
  \label{tab1}
 \end{table}
 
The possible choices of exponents, coherent with the condition on the determinant, are summarized in the following Table \ref{tab2}.
\begin{table}[ht]
\centering
\begin{tabular}{|l|l|}
\hline
  a & $\alpha \neq 0, \beta \neq 0, \gamma \neq 0$    \\
  \hline
  b    & $\alpha \neq 0, \beta \neq 0, \gamma = 0$   \\
 \hline 
  c & $\alpha = 0, \beta \neq 0, \gamma \neq 0$    \\
  \hline
  e   & $\alpha = 0, \beta = 0, \gamma \neq 0$   \\
 \hline 
  f & $\alpha \neq 0, \beta = 0, \gamma = 0$    \\
  \hline
  g    & $\alpha \neq 0, \beta = 0, \gamma \neq 0$   \\
 \hline 
\end{tabular}
 \caption{\textit{Possible coherent choices of exponents due to the determinant condition.}}
 \label{tab2}
\end{table}

These cases have to be combined with one of the two ways to made explicit the determinant condition, so, for example the c2 case means case c combined to the conditions 2 of the determinant condition. In the following we study in details the case a1 then proceeding with the other cases in a similar way. Physically, all these cases are different background space-time used to build up the gravisoliton solution.
\subsubsection*{Case a1}
Plugging the asymptotic expression into \ref{3.8} we get
\begin{equation}
\begin{aligned}
&\alpha r^{\gamma+\frac{\gamma}{2}-\frac{\delta}{2}+\alpha-1} \sim r^{n_1} \Rightarrow \frac{3 \gamma}{2}-\frac{\delta}{2}+\alpha-1=n_1; \\
&\gamma r^{\beta+\frac{\gamma}{2}-\frac{\delta}{2}+\gamma-1} \sim r^{n_2} \Rightarrow \beta + \frac{3 \gamma}{2}-\frac{\delta}{2}-1=n_2; \\
&\beta r^{\gamma+\frac{\gamma}{2}-\frac{\delta}{2}+\beta-1} \sim r^{n_3} \Rightarrow \frac{3 \gamma}{2}-\frac{\delta}{2}+\beta-1=n_3.
\label{3.11}
\end{aligned}    
\end{equation}
These equations ensure that the conserved Komar-like quantities are finite. We get the following complete matrix for our system
\begin{equation}
M|b=\begin{bmatrix}
1 & 0 & 1 & 0 & \vline & 2  \\
0 & 2 & 0 & 0  & \vline & 2  \\
1 & 0 & \frac{3}{2} & -\frac{1}{2} & \vline & n_1-1  \\
0 & 1 & \frac{3}{2} & -\frac{1}{2} & \vline & n_2-1  \\
0 & 1 & \frac{3}{2} & -\frac{1}{2} & \vline & n_3-1  \\
\end{bmatrix}
\end{equation}
From Rouché-Capelli theorem, a linear system admits solutions if only if $Rk(M)=Rk(M|b)$ where $M$ and $M|b$ are the complete and incomplete matrices associated to the linear system. In our case, both matrices have rank 4 if and only if $n_2=n_3$; moreover the solution space is a subspace of dimension $4$, hence the solution is unique. Therefore given the asymptotic data $\{n_1,n_2\}$ we get the unique radial  asymptotic expansion of the metric components fixed by the unique set of solution 
\begin{equation}
    \{\alpha=n_1-n_2+1,\beta=1,\gamma=-n_1+n_2+1,\delta=-3n_1+n_2+7\}.
\end{equation}
This solution can be expressed in terms of two parameters $\alpha$ and $\delta$ as
\begin{equation}
    \{\alpha=n_1-n_2+1,\beta=1,\gamma=-\alpha+2,\delta=-3n_1+n_2+7\};
    \label{solexp}
\end{equation}
we can note that, since $n_1,n_2 \leq 1$ we must have 
\begin{equation}
\begin{cases}
    \alpha \leq 2 & \ \ \ if \ n_2\geq 0;\\
    \alpha > 2 & \ \ \ if \ n_2< 0.\\
\end{cases}
\end{equation} 
Moreover, we get also relation between the angular functions
\begin{equation}
    f_2(\theta)=f_3(\theta)=\frac{\mathcal{BC(\theta)}}{sin^2(\theta)}\sqrt{\frac{\mathcal{C}(\theta)}{\mathcal{D}(\theta)}}, \ \ \ f_1(\theta)=\frac{\mathcal{AC(\theta)}}{sin^2(\theta)}\sqrt{\frac{\mathcal{C}(\theta)}{\mathcal{D}(\theta)}}.
\end{equation}

\subsubsection*{Case a2}
Also in the a2 case we must have $n_2=n_3$ to hava a solution;
\begin{equation}
    \{\alpha=-2n_1+8-\delta, \beta=-n_1+n_2+\alpha, \gamma=2n_1-6+\delta, \delta\}
\end{equation}
where $\delta$ is undetermined by the system. For the angular function we get
\begin{equation}
    f_2(\theta)=f_3(\theta)=0, \ \ \ f_{1}(\theta)=-\sqrt{\frac{\mathcal{C}(\theta)}{\mathcal{D}(\theta)}}.
\end{equation}
\subsubsection*{Case b1}
We have
\begin{equation}
     \{\alpha=2, \beta=1, \gamma=0, \delta=-2n_3=2(1-n_1)\};
\end{equation}
while for the angular 
\begin{equation}
    f_2(\theta)=\frac{\mathcal{BC(\theta)}}{sin^2(\theta)}\sqrt{\frac{\mathcal{C}(\theta)}{\mathcal{D}(\theta)}}, \ \ \ f_1(\theta)=\frac{\mathcal{AC(\theta)}}{sin^2(\theta)}\sqrt{\frac{\mathcal{C}(\theta)}{\mathcal{D}(\theta)}}.
\end{equation}
\subsubsection*{Case b2}
For the case b2 we have
\begin{equation}
     \{\alpha=2, \beta=n_3-n_1+2, \gamma=0, \delta=2(1-n_1)\};
\end{equation}
the angular relations are
\begin{equation}
    f_2(\theta)=0, \ \ \ f_1(\theta)=-\sqrt{\frac{\mathcal{C}(\theta)}{\mathcal{D}(\theta)}}.
\end{equation}
\subsubsection*{Case c1}
In this case we get
\begin{equation}
     \{\alpha=0, \beta=1, \gamma=2, \delta=2(3-n_2)=2(3-n_3)\};
\end{equation}
while the angular relations read
\begin{equation}
    f_2(\theta)=f_3(\theta)=\frac{\mathcal{BC}(\theta)}{sin^2(\theta)}\sqrt{\frac{\mathcal{C}(\theta)}{\mathcal{D}(\theta)}}.
\end{equation}
\subsubsection*{Case c2}
We have 
\begin{equation}
     \{\alpha=0, \beta=\delta /2 +n_2-2=\delta /2 +n_3-2, \gamma=2, \delta\},
\end{equation}
where $\delta$ is undetermined and the relations
\begin{equation}
     f_2(\theta)=f_3(\theta)=0.
\end{equation}
\subsubsection*{Case e2}
For this case we have 
\begin{equation}
    \{\alpha=0, \beta=0, \gamma=2, \delta=2(2-n_2)\},
\end{equation}
moreover 
\begin{equation}
    f_2(\theta)=0.
\end{equation}
\subsubsection*{Case f2}
We get
\begin{equation}
     \{\alpha=2, \beta=0, \gamma=0, \delta=2(1-n_1)\},
\end{equation}
and
\begin{equation}
     f_1(\theta)=-\sqrt{\frac{\mathcal{C}(\theta)}{\mathcal{D}(\theta)}}.
\end{equation}
\subsubsection*{Case g2}
For the final case we get
\begin{equation}
     \{\alpha=-2n_1+4-\delta=-n_2/5-\delta/10, \beta=0, \gamma=-\alpha+2, \delta\},
\end{equation}
where $\delta$ is undetermined and we have
\begin{equation}
     f_1(\theta)=-\sqrt{\frac{\mathcal{C}(\theta)}{\mathcal{D}(\theta)}}.
\end{equation}

\subsection{Asymptotic behavior of the $N$-solitons metric}\label{par3.1}
In order to find the asymptotic expansion of the metric \ref{metrica} we need to compute the asymptotic behaviour of \begin{equation}
\begin{aligned}
&(D)_{kl}:=(\Gamma^{-1})_{kl};\\
&(\Gamma)_{kl}:=m^{(k)}_i(g_b)_{ij}m_j^{(l)}(\rho^2+\lambda_k \lambda_l)^{-1};\\
&L_i^{(k)}:=m_j
^{(k)}(g_b)_{ji};\\
&m_i^{(k)}:=m_{2j}^{(k)}(\psi_b^{-1}(\lambda_k,\rho,z))_{ji}.
\end{aligned}    
\end{equation}
Let us start with $\psi_b^{-1}(\lambda_k,\rho,z))_{ji}$, which is determined by system  
\begin{equation}
\begin{aligned}
&\bigg(\partial_z-\frac{2\lambda^2}{\lambda^2+\rho^2}\partial_{\lambda}\bigg)\psi_b=\frac{\rho V_b-\lambda U_b}{\lambda^2+\rho^2}\psi_b,\\ 
&\bigg(\partial_{\rho}+\frac{2\lambda \rho}{\lambda^2+\rho^2}\partial_{\lambda}\bigg)\psi_b=\frac{\rho V_b+\lambda U_b}{\lambda^2+\rho^2}\psi_b;  
\label{BNMb}
\end{aligned}    
\end{equation}
we can rewrite it in spherical coordinates\footnote{Derivatives become $\frac{\partial}{\partial \rho}=\frac{\partial}{\partial r}\frac{\partial \rho}{\partial r}+\frac{\partial}{\partial \theta}\frac{\partial \rho}{\partial \theta}=sin(\theta)\frac{\partial}{\partial r}+cos(\theta)\frac{\partial}{\partial \theta}$ and $\frac{\partial}{\partial z}=\frac{\partial}{\partial r}\frac{\partial z}{\partial r}+\frac{\partial}{\partial \theta}\frac{\partial z}{\partial \theta}=cos(\theta)\frac{\partial}{\partial r}-sin(\theta)\frac{\partial}{\partial \theta}$.} 
\begin{equation}
\begin{aligned}
&\bigg(cos(\theta)\frac{\partial}{\partial r}-sin(\theta)\frac{\partial}{\partial \theta}-\frac{2\lambda^2}{\lambda^2+r^2sin^2(\theta)}\partial_{\lambda}\bigg)\psi_b=\frac{rsin(\theta) V_b-\lambda U_b}{\lambda^2+r^2sin^2(\theta)}\psi_b, \\
&\bigg(sin(\theta)\frac{\partial}{\partial r}+cos(\theta)\frac{\partial}{\partial \theta}+\frac{2\lambda \rho}{\lambda^2+r^2sin^2(\theta)}\partial_{\lambda}\bigg)\psi_b=\frac{rsin(\theta) V_b+\lambda U_b}{\lambda^2+r^2sin^2(\theta)}\psi_b;
\label{3.14}
\end{aligned}
\end{equation}
and expand them asymptotically, as done in Appendix \ref{appB}, to find a solution for the generating matrix. However the result is not handle and we prefer to follow another path. Indeed from the general theory about gravisolitons we know relations \ref{D} and \ref{D2}, which written in spherical coordinates read
\begin{equation}
    g_b(r,\theta)={\psi}_b(0,r,\theta);
\label{Ds}    
\end{equation}
therefore
\begin{equation}
 {\psi}_b(\lambda,r,\theta)=g_b(r,\theta)+\sum_{n=1}^{\infty} \lambda^n F^{(n)}_b(r,\theta)=\sum_{n=0}^{\infty} \lambda^n F_b^{(n)}(r,\theta),
\end{equation}
where $F_b^{(0)}(r,\theta)=g_b(r,\theta)$. We assume a power law expansion in the radial coordinate 
\begin{equation}
    F_b^{(n)}=\sum_{k \in \mathbb{Z}}r^kf^{(k)}(\theta) \Rightarrow  {\psi}_b(\lambda,r,\theta)=\sum_{n=0}^{\infty} \sum_{k \in \mathbb{Z}}\lambda^n r^kf^{(k)}(\theta).
    \label{expand}
\end{equation}
Now we note that due to a version of Moore-Osgood theorem we have
\begin{equation}
    \lim_{r \rightarrow \infty}\bigg(\lim_{\lambda \rightarrow 0} {\psi}_b(\lambda,r,\theta)\bigg)= \lim_{r \rightarrow \infty}g_b(r,\theta)=\lim_{\lambda \rightarrow 0}\bigg( \lim_{r \rightarrow \infty}{\psi}_b(\lambda,r,\theta)\bigg)
\label{3.18}    
\end{equation}
since the generating matrix is well-behaved, i.e. the limit for $\lambda \rightarrow 0$ converges uniformly to $g_b(r,\theta)$ and from \ref{Ds} we have $\lim_{r \rightarrow\infty}g_b(r,\theta)=\lim_{r \rightarrow \infty}{\psi}_b(0,r,\theta)$. From relation \ref{3.18} and \ref{expand} we have
\begin{equation}
\begin{aligned}
    \lim_{r \rightarrow \infty}g_b(r,\theta)&=\lim_{\lambda \rightarrow 0}\bigg( \lim_{r \rightarrow \infty}{\psi}_b(\lambda,r,\theta)\bigg)=\lim_{\lambda \rightarrow 0}\bigg( \lim_{r \rightarrow \infty}\sum_{n=0}^{\infty} \sum_{k \in \mathbb{Z}}\lambda^n r^kf^{(k)}(\theta)\bigg)=\\
    &=\lim_{\lambda \rightarrow 0}\bigg( \lim_{r \rightarrow \infty}\sum_{n=0}^{\infty} \sum_{k \in \mathbb{Z}}r^n(\pm1-cos(\theta))^n r^kf^{(k)}(\theta)\bigg)
\end{aligned}    
\end{equation}
therefore the $\lambda \rightarrow 0$ limit has to suppress all orders in $r$ which are not the leading order of the background metric; for example if the leading order for the background metric is $\mathcal{O}(r^p)$ we must have
\begin{equation}
    (\pm1-cos(\theta))^n f^{(k)}(\theta)=0 \ \ \ if \ n\neq p-k.
    \label{relcoeff}
\end{equation}
Now, if the leading order in $r$ of ${\psi}_b(\lambda,r,\theta)$ was $\lambda$-dependent, in the $\lambda \rightarrow 0$ limit the leading order would be vanishing contradicting the non-vanishing leading order, solution of system \ref{3.11}, of $g_b(r,\theta)$. Then, \ref{relcoeff} must hold also for general $\lambda$-dependence of ${\psi}_b(\lambda,r,\theta)$ and its leading order in $r$ coincide with those of $g_b(r,\theta)$. However, is a fact of life that in same cases, for example for Minkowski space-time \cite{Manzoni:2021dij}, when we compute $\psi_b(\lambda_k,r,\theta)$ the leading order of some elements of the generating matrix are different from those of the background matrix. This is a very peculiar structure due to the appearing in the the generating matrix elements of a specific combination, that is 
\begin{equation}
    r^2sin^2(\theta)-2rcos(\theta)\lambda-\lambda^2;
\end{equation}
indeed for $\lambda=\lambda_k$ this is the LHS of equation \ref{EQ2} and reduces to $-2w_k\lambda_k$. This particular feature is not captured by the general reasoning above and in order to compute the $\theta$-dependent coefficients and the subleading orders in $r$ we need to solve order by order the set of equations \ref{B.4} but this is not a simple task in a fully general setting. Case by case is easier since one knows the leading order of the background metric and can solve equations \ref{B.4} to get the leading order of the background generating matrix. We therefore introduce new factors to parametrize the leading order behavior of the background generating matrix on the poles
\begin{equation}
    (\psi_b)_{00}(\lambda_k,r,\theta) \sim r^{\epsilon_1}\mathcal{E}_1(\theta), \ \ \ (\psi_b)_{10}(\lambda_k,r,\theta) \sim r^{\epsilon_2}\mathcal{E}_2(\theta), \ \ \ (\psi_b)_{11}(\lambda_k,r,\theta) \sim r^{\epsilon_3}\mathcal{E}_3(\theta). 
\end{equation}
\\
We can now expand the interesting quantities to compute the asymptotic $N$-solitons metric. We have
\begin{equation}
\begin{aligned}
    &m_0^{(k)} \sim \frac{m_{21}^{(k)}r^{\epsilon_2}\mathcal{E}_2(\theta)-m_{20}^{(k)}r^{\epsilon_3}\mathcal{E}_3(\theta)}{r^{\epsilon_1+\epsilon_3}\mathcal{E}_1\mathcal{E}_3(\theta)-r^{2\epsilon_2}\mathcal{E}_2^2(\theta)};\\  
    &m_1^{(k)} \sim \frac{m_{21}^{(k)}r^{\epsilon_1}\mathcal{E}_1(\theta)-m_{20}^{(k)}r^{\epsilon_2}\mathcal{E}_2(\theta)}{r^{\epsilon_1+\epsilon_3}\mathcal{E}_1\mathcal{E}_3(\theta)-r^{2\epsilon_2}\mathcal{E}_2^2(\theta)},
     \label{expms}   
\end{aligned}    
\end{equation}
from which
\begin{equation}
\begin{aligned}
    &L_0^{(k)} \sim \frac{\mathcal{G}^{(k)\alpha,\beta,\beta,\alpha}_{0 (\epsilon_3,\epsilon_2,\epsilon_1\epsilon_2)}(r,\theta)}{r^{\epsilon_1+\epsilon_3}\mathcal{E}_1\mathcal{E}_3(\theta)-r^{2\epsilon_2}\mathcal{E}_2^2(\theta)};\\
     &L_1^{(k)} \sim \frac{\mathcal{G}^{(k)\beta,\gamma,\gamma,\beta}_{1 (\epsilon_3,\epsilon_2,\epsilon_1\epsilon_2)}(r,\theta)}{r^{\epsilon_1+\epsilon_3}\mathcal{E}_1\mathcal{E}_3(\theta)-r^{2\epsilon_2}\mathcal{E}_2^2(\theta)};
      \label{expLs}   
\end{aligned}    
\end{equation}
where
\begin{subequations}
\small
\begin{alignat}{2}
&\mathcal{G}^{(k)\alpha,\beta,\beta,\alpha}_{0 (\epsilon_3,\epsilon_2,\epsilon_1\epsilon_2)}(r,\theta)=m_{20}^{(k)}[r^{\epsilon_3+\alpha}\mathcal{A}\mathcal{E}_3(\theta)-r^{\epsilon_2+\beta}\mathcal{B}\mathcal{E}_2(\theta)]+m_{21}^{(k)}[r^{\epsilon_1+\beta}\mathcal{B}\mathcal{E}_1(\theta)-r^{\epsilon_2+\alpha}\mathcal{A}\mathcal{E}_2(\theta)];\\
&\mathcal{G}^{(k)\beta,\gamma,\gamma,\beta}_{1 (\epsilon_3,\epsilon_2,\epsilon_1\epsilon_2)}(r,\theta)=m_{20}^{(k)}[r^{\epsilon_3+\beta}\mathcal{B}\mathcal{E}_3(\theta)-r^{\epsilon_2+\gamma}\mathcal{C}\mathcal{E}_2(\theta)]+m_{21}^{(k)}[r^{\epsilon_1+\gamma}\mathcal{C}\mathcal{E}_1(\theta)-r^{\epsilon_2+\beta}\mathcal{B}\mathcal{E}_2(\theta)]
\end{alignat}
\end{subequations}
Using these expansions we get
\begin{equation}
\begin{aligned}
    \Gamma_{kl}\sim &\frac{{A_{kl}}+{B_{kl}}+{C_{kl}}+{D_{kl}}}{2(1\mp cos(\theta))(r^{\epsilon_1+\epsilon_3+1}\mathcal{E}_1\mathcal{E}_3(\theta)-r^{2\epsilon_2+1}\mathcal{E}^2_2(\theta))^2}.
  \label{expG}      
\end{aligned}    
\end{equation}
where
\begin{subequations}
\small
\begin{alignat}{4}
&{A_{kl}}=r^{2\epsilon_3+\alpha}\mathcal{A}(\theta)[m_{20}^{(k)}m_{20}^{(l)}\mathcal{E}^2_3(\theta)+m_{21}^{(k)}m_{21}^{(l)}\mathcal{E}^2_2(\theta)]-r^{\epsilon_2+\epsilon_3+\alpha}\mathcal{E}_2\mathcal{E}_3\mathcal{A}(\theta)[m_{20}^{(k)}m_{21}^{(l)}+m_{21}^{(k)}m_{20}^{(l)}];\\
&{B_{kl}}=-2r^{\epsilon_2+\epsilon_3+\beta}\mathcal{E}_2\mathcal{E}_3\mathcal{B}(\theta)m_{20}^{(k)}m_{20}^{(l)}-2r^{\epsilon_1+\epsilon_2+\beta}\mathcal{E}_1\mathcal{E}_2\mathcal{B}(\theta)m_{21}^{(k)}m_{21}^{(l)};\\
    &{C_{kl}}=r^{\epsilon_1+\epsilon_3+\beta}\mathcal{E}_1\mathcal{E}_3\mathcal{B}(\theta)[m_{20}^{(k)}m_{21}^{(l)}+m_{21}^{(k)}m_{20}^{(l)}]+r^{2\epsilon_2+\beta}\mathcal{E}^2_2\mathcal{B}(\theta)[m_{20}^{(k)}m_{21}^{(l)}+m_{21}^{(k)}m_{20}^{(l)}];\\
    &{D_{kl}}=r^{2\epsilon_2+\gamma}\mathcal{C}(\theta)[m_{20}^{(k)}m_{20}^{(l)}\mathcal{E}^2_2(\theta)+m_{21}^{(k)}m_{21}^{(l)}\mathcal{E}^2_1(\theta)]-r^{\epsilon_1+\epsilon_2+\gamma}\mathcal{E}_1\mathcal{E}_2\mathcal{C}(\theta)[m_{20}^{(k)}m_{21}^{(l)}+m_{21}^{(k)}m_{20}^{(l)}].
\end{alignat}    
\end{subequations} 
To find the asymptotic behaviour of $(D)_{kl}$ we write it as
\begin{equation}
    (D)_{kl}=(\Gamma^{-1})_{kl}=\frac{(Adj(\Gamma))_{kl}}{\det(\Gamma)}
\end{equation}
and we have to taken into account that for an $N$-solitons solution $(\Gamma)_{kl}$ is an $N \times N$ matrix. The determinant is given by
\begin{equation}
    {\displaystyle \det(\Gamma)=\sum _{\sigma \in S_{N}}\left(\operatorname {sgn}(\sigma )\prod _{s=0}^{N}(\Gamma)_{s,\sigma _{s}}\right)};
\end{equation}
asymptotically, the generic addend is
\begin{equation}
\begin{aligned}
    \operatorname {sgn}(\sigma )\prod _{s=0}^{N}(\Gamma)_{s,\sigma _{s}} &\sim \operatorname {sgn}(\sigma ) \bigg(\frac{1}{2(1\mp cos(\theta))(r^{\epsilon_1+\epsilon_3+1}\mathcal{E}_1\mathcal{E}_3(\theta)-r^{2\epsilon_2+1}\mathcal{E}^2_2(\theta))^2}\bigg)^N \times \\
    & \times \prod _{s=0}^{N} [{A_{s\sigma_s}}+{B_{s\sigma_s}}+{C_{s\sigma_s}}+{D_{s\sigma_s}}].
\end{aligned}    
\end{equation}
Using the expression above for the determinant we can also compute the asymptotic behaviour of the cofactor which is computed by a determinant of an $(N-1) \times (N-1)$ matrix; therefore we compute that
\begin{equation}
\begin{aligned}
    (D)_{kl} &\sim [{2(1\mp cos(\theta))(r^{\epsilon_1+\epsilon_3+1}\mathcal{E}_1\mathcal{E}_3(\theta)-r^{2\epsilon_2+1}\mathcal{E}^2_2(\theta))^2}] \times \\
    &\times \frac{\sum _{\tilde{\sigma} \in S_{N-1}}\operatorname {sgn}(\tilde{\sigma} )\prod _{s=0}^{N-1} [{A_{s\tilde{\sigma}_s}}+{B_{s\tilde{\sigma}_s}}+{C_{s\tilde{\sigma}_s}}+{D_{s\tilde{\sigma}_s}}]}{\sum _{\sigma \in S_{N}}\operatorname {sgn}({\sigma})\prod _{s=0}^{N} [{A_{s\sigma_s}}+{B_{s\sigma_s}}+{C_{s\sigma_s}}+{D_{s\sigma_s}}]};
 \label{expD}   
\end{aligned}    
\end{equation}
where $s \neq k$ and $\tilde{\sigma}_s \neq l$. One may wonder and questioning about the effective usefulness and easy handling of expansions \ref{expms}, \ref{expLs}, \ref{expG} and \ref{expD}; the point is that they are completely general but, case by case, one knows exponents $\{\alpha,\beta,\gamma,\epsilon_1,\epsilon_2,\epsilon_3\}$ and angular functions $\{\mathcal{A}(\theta),\mathcal{B}(\theta),\mathcal{C}(\theta),\mathcal{E}_1(\theta),\mathcal{E}_2(\theta),\mathcal{E}_3(\theta)\}$ and can extract the leading orders in a simple way. 

We can finally expand the $N$-solitons metric; let us start with the combination appearing in the $g_{ij}$ elements, for example in $g_{00}$
\begin{subequations}
\small
\begin{equation}
\begin{aligned}
&\sum_{k,l=1}^N(D)_{kl}\lambda_k^{-1}\lambda_l^{-1}L_0^{(k)}L_0^{(l)} \sim \\
&\sim \sum_{k,l=1}^N K  \mathcal{G}^{(k)\alpha,\beta,\beta,\alpha}_{0 (\epsilon_3,\epsilon_2,\epsilon_1\epsilon_2)}(r,\theta)\mathcal{G}^{(l)\alpha,\beta,\beta,\alpha}_{0 (\epsilon_3,\epsilon_2,\epsilon_1\epsilon_2)}(r,\theta)\frac{\sum _{\tilde{\sigma} \in S_{N-1}}\operatorname {sgn}(\tilde{\sigma} )\prod _{s=0}^{N-1} [{A_{s\tilde{\sigma}}}+{B_{s\tilde{\sigma}}}+{C_{s\tilde{\sigma}}}+{D_{s\tilde{\sigma}}}]}{\sum _{\sigma \in S_{N}}\operatorname {sgn}({\sigma})\prod _{s=0}^{N} [{A_{s\tilde{\sigma}_s}}+{B_{s\tilde{\sigma}_s}}+{C_{s\tilde{\sigma}_s}}+{D_{s\tilde{\sigma}_s}}]},
\end{aligned}
\end{equation}
\end{subequations}
where $K={2\frac{(1\mp cos(\theta))}{(\pm 1 -cos(\theta))^2}}=\frac{2}{(1\mp cos(\theta))}$ since $(\pm 1 -cos(\theta))^2=( 1 \mp cos(\theta))^2$. Similar expansions hold for $g_{01}=g_{10}$ and $g_{11}$.
Putting it all together, the metric component on the $N$-gravisoliton have the following asymptotic expansion 
\begin{equation}
\begin{aligned}
&g_{22}^{(ph)} \sim \mathcal{M}_{N}(\theta)\bigg(\frac{1}{(r^{\epsilon_1+\epsilon_3+1}\mathcal{E}_1\mathcal{E}_3(\theta)-r^{2\epsilon_2+1}\mathcal{E}^2_2(\theta))^2}\bigg)^N r^{\delta-\frac{N^2}{2}+2N}  \times \\
& \ \ \ \ \ \ \ \ \ \times
\sum _{\sigma \in S_{N}}\big(\operatorname {sgn}(\sigma)  \prod _{s=0}^{N}[{A_{s\sigma_s}}+{B_{s\sigma_s}}+{C_{s\sigma_s}}+{D_{s\sigma_s}}]\big);\\
&g_{00}^{(ph)}\sim \pm \mathcal{N}_N(\theta) \bigg[r^{\alpha}\mathcal{A}(\theta)-\sum_{k,l=1}^N K \mathcal{K} \mathcal{G}^{(k)\alpha,\beta,\beta,\alpha}_{0 (\epsilon_3,\epsilon_2,\epsilon_1\epsilon_2)}(r,\theta)\mathcal{G}^{(l)\alpha,\beta,\beta,\alpha}_{0 (\epsilon_3,\epsilon_2,\epsilon_1\epsilon_2)}(r,\theta)\bigg];\\
&g_{11}^{(ph)}\sim \pm \mathcal{N}_N(\theta) \bigg[r^{\gamma}\mathcal{C}(\theta)-\sum_{k,l=1}^N K\mathcal{K}\mathcal{G}^{(k)\alpha,\beta,\beta,\alpha}_{1 (\epsilon_3,\epsilon_2,\epsilon_1\epsilon_2)}(r,\theta)\mathcal{G}^{(l)\beta,\gamma,\gamma,\beta}_{1 (\epsilon_3,\epsilon_2,\epsilon_1\epsilon_2)}(r,\theta)\bigg]; \\
&g_{10}^{(ph)}=g_{01}^{(ph)}\sim \pm \mathcal{N}_N(\theta) \bigg[r^{\beta}\mathcal{B}(\theta)-\sum_{k,l=1}^N K\mathcal{K}\mathcal{G}^{(k)\alpha,\beta,\beta,\alpha}_{0 (\epsilon_3,\epsilon_2,\epsilon_1\epsilon_2)}(r,\theta)\mathcal{G}^{(l)\beta,\gamma,\gamma,\beta}_{1 (\epsilon_3,\epsilon_2,\epsilon_1\epsilon_2)}(r,\theta)\bigg];
\label{metricaespansa}    
\end{aligned}    
\end{equation}
where
\begin{equation}
\begin{aligned}
    &\mathcal{M}_{N}(\theta)= 4^{N(N-1)+2}C\mathcal{D}(\theta)[(\pm1- cos(\theta))]^{N^2+N}sin(\theta)^{-\frac{N^2}{2}}(2\mp 2cos(\theta))^{-N};\\
    &\mathcal{K}=\frac{\sum _{\tilde{\sigma} \in S_{N-1}}\operatorname {sgn}(\tilde{\sigma} )\prod _{s=0}^{N-1} [{A_{s\tilde{\sigma}_s}}+{B_{s\tilde{\sigma}_s}}+{C_{s\tilde{\sigma}_s}}+{D_{s\tilde{\sigma}_s}}]}{\sum _{\sigma \in S_{N}}\operatorname {sgn}({\sigma})\prod _{s=0}^{N} [{A_{s\sigma_s}}+{B_{s\sigma_s}}+{C_{s\sigma_s}}+{D_{s\sigma_s}}]};\\
    &\mathcal{N}_N=(\theta)\bigg[\frac{(\pm 1-cos(\theta))}{sin(\theta)}\bigg]^N.
    \label{const}
\end{aligned}    
\end{equation}
We note that for the $g_{22}^{(ph)}$ component we have more choices for the asymptotic expansion according to the relative signs in front of the square roots in the definition of the poles due to the presence of the factors $\frac{1}{(\lambda_k-\lambda_l)^2}$ which give $\frac{1}{4r^2}$ if the signs disagree while give $\frac{1}{w_k-w_l}$ if signs agree. Therefore the most general expansion for $g_{22}^{(ph)}$ is
\begin{equation}
\begin{aligned}
&g_{22}^{(ph)} \sim \mathcal{M}_{N}(\theta)\bigg(\frac{1}{(r^{\epsilon_1+\epsilon_3+1}\mathcal{E}_1\mathcal{E}_3(\theta)-r^{2\epsilon_2+1}\mathcal{E}^2_2(\theta))^2}\bigg)^N r^{\delta-\frac{N^2}{2}+N(N+1)-N(N-1)+2\#} \times \\
& \ \ \ \ \ \ \ \ \ \times \bigg(\prod_{j=1}^{\#}\frac{1}{(w_k-w_l)_j^2}\bigg)
\sum _{\sigma \in S_{N}}\big(\operatorname {sgn}(\sigma)  \prod _{s=0}^{N}[{A_{s\sigma_s}}+{B_{s\sigma_s}}+{C_{s\sigma_s}}+{D_{s\sigma_s}}]\big);\\
\label{metricaespansag22gen}    
\end{aligned}    
\end{equation}
where
\begin{equation}
\begin{aligned}
    &\mathcal{M}_{N}(\theta)= 4^{N(N-1)-2\#+2}C\mathcal{D}(\theta)[(\pm1- cos(\theta))]^{N^2+N}sin(\theta)^{-\frac{N^2}{2}}(2\mp 2cos(\theta))^{-N},
    \label{const1}
\end{aligned}    
\end{equation}
and  $\#$ is the number of poles couple with concordant root signs.

We can check that choosing Minkowski metric as background and computing the asymptotic behaviour of the background generating matrix, the asymptotic behaviour of $m_0^{(k)}, m_1^{(k)},L_0^{(k)},l_1^{(k)}, \Gamma_{kl}, D_{kl}$ and of the metric component is exactly what was expected \cite{Belinski:2001ph},\cite{Manzoni:2021dij}. We note that the physical information of how many solitons make up the solution is reached by all the metric components meaning again that non-diagonal solution can be generated even if the background solution was diagonal; this is the case of Kerr black hole $2$-solitons solution. Moreover, there are no reasons why the asymptotic behavior of the chosen background solution must be preserved by the $N$-soliton solution and in general this is no the case. Indeed, let us consider the following example. Consider a 
background metric with $\delta=1$ and with $\epsilon_1=\epsilon_2=\epsilon_3=0$. This is no restrictive since the determination of the asymptotic behavior of the background generating matrix does not depend on $\delta$. In this case the $N$-solitons solution give us a $g_{22}^{(ph)}$ metric component with 
asymptotic behavior $r^{1-\frac{N^2}{2}}r^{h(\alpha,\beta,\gamma)}$ where $h(\alpha,\beta,\gamma) \in \frac{1}{2}\mathbb{Z}$ has to derived case by case depending on $\alpha,\beta$ and $\gamma$ but then it is fixed. However, surely there exist a $N$ for which $1-\frac{N^2}{2}+h(\alpha,\beta,\gamma)\neq 0$. This show that the $N$-solitons solution has no to preserve the asymptotic behavior of the background metric in general, at least in the assumption of Komar-like quantities are finite. Therefore it is a well posed question ask for asymptotic symmetries with expectation that these may depend on the number of solitons that make up the solution.
\subsection{Leading order of the asymptotic symmetries and the UCS group}\label{subpar3.2}
In order to discuss asymptotic symmetries let us find the asymptotic Killing vectors solutions of the asymptotic Killing equation, i.e. Killing equation for the asymptotic $N$-solitons metric.  As known, a vector field $\xi$ is a Killing field if the Lie derivative with respect to $\xi$ of the metric vanishes
\begin{equation}
    {\mathcal  {L}}_{{\xi}}g=0,
\end{equation}
in terms of the Levi-Civita connection, we can write it as
\begin{equation}
    {\displaystyle g\left(\nabla _{Y}\xi,Z\right)+g\left(Y,\nabla _{Z}\xi\right)=0,}
\end{equation}
for all vectors $Y$ and $Z$. Moreover in local coordinates expressed by a coordinate fields, Killing equation assume its standard form
\begin{equation}
    {\displaystyle \nabla _{\mu }\xi_{\nu }+\nabla _{\nu }\xi_{\mu }=\xi_{(\mu ; \nu)}=\partial_{\mu}\xi_{\nu}+\partial_{\nu}\xi_{\mu}-2\Gamma^{\rho}_{\mu \nu}\xi_{\rho}=0};
\end{equation}
moreover, we assume a homogeneous expansion for the Killing vectors of the form
\begin{equation}
    \xi_{\mu}=\sum_{l \in \mathbb{Z}} \frac{\bar{\xi}_{\mu}^{(l)}(\theta)}{r^l} \ \ \ \ \ \mu=0,1,2,3.
\end{equation}
Our goal is to determine the first possible $l$ of the expansion above; this information can be obtained looking at the Killing equation and requiring that it is satisfied only asymptotically, i.e. it is not really necessary for it to be exactly zero as long as it is satisfied within the limit of large radial distances. We focus on the Killing equation with $(\mu, \nu)=(3,3)$ this reduces, using results in \ref{cri}, to
\begin{equation}
    2\partial_3\xi_3-2\Gamma_{33}^3\xi_3 \sim \mathcal{O}(r^0).
\end{equation}
For $N=0$ we have, inserting the Killing vector expansion 
\begin{equation}
    2\sum_{l \in \mathbb{Z}} \frac{\bar{\xi}_{3}^{'(l)}(\theta)}{r^l}-\frac{\mathcal{D}'(\theta)}{\mathcal{D}(\theta)}\sum_{l \in \mathbb{Z}} \frac{\bar{\xi}_{3}^{(l)}(\theta)}{r^l}\sim \mathcal{O}(r^0),
\end{equation}
which means that the leading order is $l=0$ since for $l>0$ we have subleading terms while for $l<0$ we have overleading terms; moreover coefficients must satisfy the equation
\begin{equation}
    2\bar{\xi}_{3}^{'(l)}(\theta)-\frac{\mathcal{D}'(\theta)}{\mathcal{D}(\theta)}\bar{\xi}_{3}^{(l)}(\theta)=0 \ \ \ \ \ \forall l \in (-\infty,0],
\end{equation}
Let us continue with the case of $N=2$; in this case using the Christoffel symbols derived in \ref{gamman2} the equation is
\begin{equation}
    2\sum_{l \in \mathbb{Z}} \frac{\bar{\xi}_{3}^{'(l)}(\theta)}{r^l}-2\big(\Gamma^{3}_{33}\big)_{N=2}\sum_{l \in \mathbb{Z}} \frac{\bar{\xi}_{3}^{(l)}(\theta)}{r^l}\sim \mathcal{O}(r^0),
\end{equation}
since $\big(\Gamma^{3}_{33}\big)_{N=2}$ turn out to has a leading order independent of $r$ then, again, the leading order of the Killing vector is $l=0$; moreover coefficients satisfy
\begin{equation}
    2\bar{\xi}_{3}^{'(l)}(\theta)-2\big(\Gamma^{3}_{33}\big)_{N=2}\bar{\xi}_{3}^{(l)}(\theta)=0 \ \ \ \ \ \forall l \in (-\infty,0].
\end{equation}
The same kind of analysis can be done in the general $N$ case but we note that despite the leading order is the same for all $N$ the angular dependence can be different; this can be seen since the Christoffel symbols have different angular dependence depending on $N$ since the $N$-solitons metric has. Therefore, in the case of Killing vectors whose components fall-off with the same behaviour, the leading order is $r^0$ while the specific case coefficients have to be derived case by case but their angular dependence depend from the number of solitons. The other component of the Killing equations give us relations between expansion parameters, i.e. the exponents of $r$ in the metric expansion, and $l$; knowing that $l=0$ some of these relations may not be satisfied and we find that not for all choices of expansion parameters we have a non-vanishing leading order. In these cases the Killing vectors decay too quickly and do not generate any symmetry algebra on the corner at infinite distance.

The point now is to compute the algebra by investigating the vector field bracket in order to make a link with the corner proposal. The corner proposal \cite{ciambelli2023asymptotic} is based on the fact that gauge symmetries are classical redundancies of the system that are not expected to survive in quantum gravity. This applies in particular to diffeomorphisms that are pure gauge; however, there are diffeomorphisms that are asymptotic symmetries which are not redundancy, they physically act on the field space. The corner proposal focuses on these symmetries, and posits that they survive in quantum gravity. The central point is that a gravitational theory is described by a set of charges and their algebra at corners.

The only Killing vectors that can generate a symmetry algebra on the infinite distance corner are the leading order ones; we refer to them sympli as $\xi^{\mu}$. We can expand a vector field in the coordinate basis on the corner obtained in the radial limit at fixed time $\partial_i$ with $i=1,3$ and the "normal" coordinates $\partial_a$ with $a=0,2$. Therefore 
\begin{equation}
\xi^{\mu}\partial_{\mu}=\xi^i\partial_i+\xi^a\partial_a.
\end{equation}
It is ease to see that these Killing vectors generate, as expected, a subalgebra of the \textfrak{ucs} 
algebra and therefore its exponential generate a subgroup of the UCS group. This is because the Lie bracket gives us, where expressed in coordinates
\begin{equation}
    [\xi^i\partial_i+\xi^a\partial_a, \eta^i\partial_i+\eta^a\partial_a]=[\xi^i\partial_i, \eta^i\partial_i]^j\partial_j+(\xi^i\partial_i\eta^a-\eta^i\partial_i\xi^a)\partial_a;
\end{equation}
the first term is $\textfrak{Diff}(S)$ while the second one is the action of $\textfrak{Diff}(S)$ on $\mathbb{R}^2$ where $S$ is the infinite distance corner whose coordinates are the angles. Therefore the axialgravisoliton corner symmetry algebra (\textfrak{agcsa}) generated by the leading order Killing vectors is 
\begin{equation}
    \textfrak{agcsa}=  \textfrak{Diff}(S) \oplus_{sd} \mathbb{R}^2.
\end{equation}
where $\oplus_{sd}$ stands for the semi-direct sum of Lie algebras.

\section{Conclusion}
Gravitational solitons or gravisolitons are particularly exact solutions of the Einstein gravity that can be extended to Einstein-Maxwell theory \cite{1978ZhETF..75.1953B, 1980JETPL..32..277A}. The theory of gravitational solitons is essentially linked to the Inverse Scattering Method (ISM) and many of the most useful metrics used, like those of black holes, are nothing but gravisolitons \cite{Belinski:2001ph, Manzoni:2021dij}. Indeed, gravisolitons incorporate many of physically relevant solutions such as black holes and cosmological one \cite{Belinski:2001ph} but also more exotic recently proposed solutions \cite{Vigano:2022hrg}. Moreover gravitational solitons are a good candidate of dark matters since they have the right properties and the right characteristics \cite{Asselmeyer_Maluga_2020}; nevertheless, they are fundamental for the completeness hypothesis thanks to the charges they carry \cite{mcnamara2021gravitational}. Despite this, the fundamental role and implication of gravisolitons is not yet fully understood; a systematic formal and mathematical study of them can be useful to understand how and when gravisolitons enter in the game both classically and quantumly; shedding light on this fundamental but not fully explored sector of pure gravity. The formal and mathematical study of gravisolitons could be useful also in string theory since several gravitational string backgrounds can be interpreted as soliton solutions \cite{Bakas:1996gs}.
\subsection{Results and discussion}
In this work we study some formal and mathematical properties of a particular class of gravisolitons: axialgravisolitons. First of all we study the background metric compatible with the requests on the determinant of the metric, which is an essential condition to have physical metrics, and with the finiteness of Komar-like integrals associated, which are computed in Appendix $\ref{appA}$. These conditions force us to divide the asymptotic expansion of the background solution in cases summarized in Table \ref{tab2}. According to each case, the set of exponents $\{\alpha,\beta,\gamma,\delta\}$ is different but respects both the determinant condition and the Komar-like integrals finitness. Giving only these conditions on the background solution we find the expansion for any value of the soliton numbers $N$. According to the parameters $\{\alpha,\beta,\gamma,\delta,\epsilon_1,\epsilon_2,\epsilon_3\}$, the final $N$-soliton metric can develop different asymptotic behaviour with respect the background solution used to generate the solitons. This is reported in \ref{metricaespansa}, \ref{const},  \ref{metricaespansag22gen} and \ref{const1}. As technical note we underline that the asymptotic expansion of the component $g_{22}^{(ph)}$ depends strongly on the relative signs in front of the square roots in the definition of the poles. The asymptotic expansion developed can play, for $N$-axialgravisoliton space-time, a similar role of Bondi expansion for asymptotically flat space-time and Fefferman-Graham expansion for asymototically AdS space-time. We showed at the end of Paragraph \ref{par3.1} that, in general, the $N$-soliton metric has no to preserve the asymptotically behaviour of the background metric even developing off-diagonal terms as in the case of Minkowski background and Kerr black hole (which is a 2-pole axialgravisoliton); therefore it
is a well posed question ask for asymptotic symmetries with the expectation that these
may depend on the number of solitons that make up the solution. In order to compute the asymptotic Killing vectors we use the expansion derived to compute its Christoffel symbols in Appendix $\ref{appB}$. This is in fact the case since the leading order coefficients satisfy different differential equations according to the number of solitons $N$, however the radial leading order turn out to be the same in every case. The algebra generated by these Killing vectors can be computed looking to their bracket and turn out to be a subalgebra of the universal corner symmetry algebra given by $\textfrak{agcsa}=  \textfrak{Diff}(S) \oplus_{sd} \mathbb{R}^2$. This suggests a positive feedback for the corner proposal which is explicitly tested and open the way to a possible quantization to the non-asymptotically flat sector of gravity by studying the representations of \textfrak{agcsa} since $N$-axialgravisoliton has no to be necessary asymptotically flat. Moreover, for the sake of speculation, \textfrak{agcsa} could play a similar role of the conformal algebra for AdS space-time or the  \textfrak{BMS} algebra for asymptotically flat space-time upon conformal compactification: we could construct a QFT at the boundary of the $N$-axialgravisoliton solution with symmetry algebra given by \textfrak{agcsa} and this could be dual to some gravitational process which take place in the bulk of the $N$-axialgravisoliton metric. This is a kind of holographic correspondence we can call $N$-axialgravisoliton/\textfrak{agcsa}-QFT correspondence. The applicability and construction of this correspondence can be material for future work with the aims to better understand the role of gravisolitons in classical and quantum gravity. \\
Moreover, in the spirit of Strominger IR triangle \cite{strominger2018lectures}, \textfrak{agcsa} is only one of the three corners. These axialgravisoliton asymptotic symmetries should be related, on the one hand, to some version or extension of the soft theorem on curved background\footnote{It is well-known that a global description of the S-matrix may not exist in an arbitrary curved space-time. In \cite{Mandal:2019bdu}, authors give a local construction of S-matrix in quantum field theory in curved space-time using Riemann-normal coordinates which mimics the methods, generally used in Minkowski spacetime.} and, on the other hand, to an observable memory effect\footnote{To distinguish this effect from the gravitational and spin-gravitational memory effects \cite{Pasterski_2016, Strominger:2014pwa} we call it axialgravisoliton memory effect.}.

\begin{figure}[H]
\centering
\tikzset{every picture/.style={line width=0.75pt}} %set default line width to 0.75pt        

\begin{tikzpicture}[x=0.75pt,y=0.75pt,yscale=-1,xscale=1]
%uncomment if require: \path (0,235); %set diagram left start at 0, and has height of 235

%Flowchart: Extract [id:dp08192048462199608] 
\draw   (332.5,28.5) -- (433,180) -- (232,180) -- cycle ;
%Shape: Circle [id:dp738434772680189] 
\draw  [color={rgb, 255:red, 0; green, 0; blue, 0 }  ,draw opacity=1 ][fill={rgb, 255:red, 208; green, 2; blue, 27 }  ,fill opacity=1 ] (329.25,28.5) .. controls (329.25,26.71) and (330.71,25.25) .. (332.5,25.25) .. controls (334.29,25.25) and (335.75,26.71) .. (335.75,28.5) .. controls (335.75,30.29) and (334.29,31.75) .. (332.5,31.75) .. controls (330.71,31.75) and (329.25,30.29) .. (329.25,28.5) -- cycle ;
%Shape: Circle [id:dp5711972004640214] 
\draw  [color={rgb, 255:red, 0; green, 0; blue, 0 }  ,draw opacity=1 ][fill={rgb, 255:red, 74; green, 144; blue, 226 }  ,fill opacity=1 ] (228.75,180) .. controls (228.75,178.21) and (230.21,176.75) .. (232,176.75) .. controls (233.79,176.75) and (235.25,178.21) .. (235.25,180) .. controls (235.25,181.79) and (233.79,183.25) .. (232,183.25) .. controls (230.21,183.25) and (228.75,181.79) .. (228.75,180) -- cycle ;
%Shape: Circle [id:dp05956709470925925] 
\draw  [color={rgb, 255:red, 0; green, 0; blue, 0 }  ,draw opacity=1 ][fill={rgb, 255:red, 189; green, 16; blue, 224 }  ,fill opacity=1 ] (429.75,180) .. controls (429.75,178.21) and (431.21,176.75) .. (433,176.75) .. controls (434.79,176.75) and (436.25,178.21) .. (436.25,180) .. controls (436.25,181.79) and (434.79,183.25) .. (433,183.25) .. controls (431.21,183.25) and (429.75,181.79) .. (429.75,180) -- cycle ;

% Text Node
\draw (445,161.4) node [anchor=north west][inner sep=0.75pt]  [color={rgb, 255:red, 189; green, 16; blue, 224 }  ,opacity=1 ]  {$ \begin{array}{l}
axialgravisoliton\\
\ memory\ effect
\end{array}$};
% Text Node
\draw (196,3.4) node [anchor=north west][inner sep=0.75pt]  [color={rgb, 255:red, 208; green, 2; blue, 27 }  ,opacity=1 ]  {$soft\ theorem\ on\ axialgravisoliton\ background$};
% Text Node
\draw (89,163.4) node [anchor=north west][inner sep=0.75pt]  [color={rgb, 255:red, 74; green, 144; blue, 226 }  ,opacity=1 ]  {$ \begin{array}{l}
axialgravisoliton\\
\ asymptotic\ symmetry\ 
\end{array}$};
% Text Node
\draw (272,123.4) node [anchor=north west][inner sep=0.75pt]    {$ \begin{array}{l}
axialgravisoliton\ \\
IR\ triangle
\end{array}$};

\end{tikzpicture}
\caption{Schematization of the axialgravisoliton IR triangle.}
\end{figure}

Beyond the technical observational issues, the possible observation of this axialgravisoliton memory effect is complicated by the fact that we should be in the asymptotic region of a space-time with a certain number of axialsymmetric gravisolitons. The study and characterization of these other two corners of the IR triangle may be attacked in the near future.\\
As other future applications and extensions of this work, it would be interesting to consider the case of higher derivative and $f(R)$ gravity. Both these cases are natural evolutions of Einstein gravity at UV regimes and the understanding, in these framework, of the gravisolitons role, of their properties and asymptotic symmetries could help in the road to quantum gravity. Other important cases of extension are those with spinors and supersymmetry, i.e. Einstein-spinors gravity and SUGRA theories. However, in all such contexts, the theory of gravisolitons, in the sense of ISM and Darboux transformation, is poorly developed. A first step in considering these extensions would be to develop a systematic metric valid for every number of solitons such as \ref{metrica}. \\

\subsection*{Acknowledgment}
I want to to thanks Paolo Maria Santini who introduced me, at the time of my master's degree, to non-linear waves and gravitational solitons. I thank my other half, Maria Luisa Limongi for having, once again, supported my work by also helping me when some computations were struggling to move forward. I thank Luca Ciambelli for the interesting seminar with which he introduced me to the corner proposal; I also thank him for interesting discussions on the matter. Finally, I thank my PhD advisor Dario Francia.

\appendix

\section{Evaluation of Komar-like quantities}\label{appA}
In an explicit way metric and its inverse read
\begin{equation}
(g_b)_{\mu \nu}\begin{bmatrix}
A & B & 0 & 0\\
B & C & 0 & 0\\
0 & 0 & (g_b)_{rr} & 0\\
0 & 0 & 0 & r^2(g_b)_{rr}
\end{bmatrix}, \ \ \ \ \ (g_b)^{\mu \nu}\begin{bmatrix}
-\frac{C}{r^2sin^2(\theta)} & \frac{B}{r^2sin^2(\theta)} & 0 & 0\\
\frac{B}{r^2sin^2(\theta)} & -\frac{A}{r^2sin^2(\theta)} & 0 & 0\\
0 & 0 & \frac{1}{(g_b)_{rr}} & 0\\
0 & 0 & 0 & \frac{1}{r^2(g_b)_{rr}}.
\end{bmatrix}
\end{equation}
Let us start with the first Komar-like integral, we need $*dk$. $k^{\mu}=(1,0,0,0)$ and so $k_{\mu}=(g_b)_{\mu \nu}k^{\nu}=(g_b)_{\mu t}$; the form associated is $k=(g_b)_{\mu t}dx^{\mu}=Adt+Bd\phi$ and its exterior derivative reads
\begin{equation}
    dk=\frac{\partial A}{\partial r}dr \wedge dt+\frac{\partial A}{\partial \theta}d\theta \wedge dt+\frac{\partial B}{\partial r}dr \wedge d\phi+\frac{\partial B}{\partial \theta}d\theta \wedge d\phi.
\end{equation}
Taking the Hodge dual we get
\begin{subequations}
\small
\begin{equation}
\begin{aligned}
    &(*dk)_{\mu \nu}=\frac{\epsilon_{\mu \nu \rho \sigma}}{2}(g_b)^{\rho \rho'}(g_b)^{\sigma \sigma'}(dk)_{\rho'\sigma'}=\\
    &=\frac{\epsilon_{\mu \nu \rho \sigma}}{2}[(g_b)^{\rho r}(g_b)^{\sigma t}(dk)_{rt}+(g_b)^{\rho \theta}(g_b)^{\sigma t}(dk)_{\theta t}+(g_b)^{\rho r}(g_b)^{\sigma \phi}(dk)_{r\phi}+(g_b)^{\rho \theta}(g_b)^{\sigma \phi}(dk)_{\theta \phi}].
\end{aligned}    
\end{equation}
\end{subequations}
A similar procedure can be done for computing $*dm$: $m=Bdt+Cd\phi$ and the final result is
\begin{subequations}
\small
\begin{equation}
\begin{aligned}
    &(*dm)_{\mu \nu}=\frac{\epsilon_{\mu \nu \rho \sigma}}{2}(g_b)^{\rho \rho'}(g_b)^{\sigma \sigma'}(dm)_{\rho'\sigma'}=\\
    &=\frac{\epsilon_{\mu \nu \rho \sigma}}{2}[(g_b)^{\rho r}(g_b)^{\sigma t}(dm)_{rt}+(g_b)^{\rho \theta}(g_b)^{\sigma t}(dm)_{\theta t}+(g_b)^{\rho r}(g_b)^{\sigma \phi}(dm)_{r\phi}+(g_b)^{\rho \theta}(g_b)^{\sigma \phi}(dm)_{\theta \phi}],
\end{aligned}    
\end{equation}
\end{subequations}
with 
\begin{equation}
    dm=\frac{\partial B}{\partial r}dr \wedge dt+\frac{\partial B}{\partial \theta}d\theta \wedge dt+\frac{\partial C}{\partial r}dr \wedge d\phi+\frac{\partial C}{\partial \theta}d\theta \wedge d\phi.
\end{equation}

To calculate the Komar quantities $K_t$ and $K_{\phi}$ one needs to choose an appropriate boundary surface. As underlined in \cite{1984JMP....25..992C} and \cite{Modak_2011} a good choice is the boundary of a spatial three volume characterised by a
constant $r$ and $dt=-\frac{A}{B}d\phi$; In this way, infinitesimally close points are simultaneous events. Since $r$ is constant the dual Killing forms reduced to
\begin{equation}
\begin{aligned}
    &(*dk)_{\mu \nu}=\frac{1}{2}\epsilon_{\mu \nu \rho \sigma}[(g_b)^{\rho r}(g_b)^{\sigma t}(dk)_{rt}+(g_b)^{\rho r}(g_b)^{\sigma \phi}(dk)_{r\phi}]=\\
    &=\frac{(g_b)^{rr}}{2}\{(dk)_{rt}[\epsilon_{\mu \nu r t}(g_b)^{tt}+\epsilon_{\mu \nu r \phi}(g_b)^{\phi t}]+(dk)_{r\phi}[\epsilon_{\mu \nu r t}(g_b)^{t\phi}+\epsilon_{\mu \nu r \phi}(g_b)^{\phi \phi}]\}; \\
    &(*dm)_{\mu \nu}=\frac{1}{2}\epsilon_{\mu \nu \rho \sigma}[(g_b)^{\rho r}(g_b)^{\sigma t}(dm)_{rt}+(g_b)^{\rho r}(g_b)^{\sigma \phi}(dm)_{r\phi}]=\\
    &=\frac{(g_b)^{rr}}{2}\{(dm)_{rt}[\epsilon_{\mu \nu r t}(g_b)^{tt}+\epsilon_{\mu \nu r \phi}(g_b)^{\phi t}]+(dm)_{r\phi}[\epsilon_{\mu \nu r t}(g_b)^{t\phi}+\epsilon_{\mu \nu r \phi}(g_b)^{\phi \phi}]\}
\end{aligned}    
\end{equation}
and the only non-vanishing elements are
\begin{equation}
\begin{aligned}
&(*dk)_{\theta \phi}=\frac{(g_b)^{rr}}{2}\epsilon_{\theta \phi r t }[(g_b)^{tt}(dk)_{rt}+(g_b)^{t\phi}(dk)_{r\phi}];\\
&(*dk)_{\theta t}=\frac{(g_b)^{rr}}{2}\epsilon_{\theta t r \phi }[(g_b)^{\phi t}(dk)_{rt}+(g_b)^{\phi  \phi}(dk)_{r\phi}],
\end{aligned}    
\end{equation}
for $*dk$ and 
\begin{equation}
\begin{aligned}
&(*dm)_{\theta \phi}=\frac{(g_b)^{rr}}{2}\epsilon_{\theta \phi r t }[(g_b)^{tt}(dm)_{rt}+(g_b)^{t\phi}(dm)_{r\phi}];\\
&(*dm)_{\theta t}=\frac{(g_b)^{rr}}{2}\epsilon_{\theta t r \phi }[(g_b)^{\phi t}(dm)_{rt}+(g_b)^{\phi  \phi}(dm)_{r\phi}],
\end{aligned}    
\end{equation}
for $*dm$.

Since we are performing an integral over a over a surface of simultaneous events terms $(*dk)_{\theta t}$ and $(*dm)_{\theta t}$ do not have to be included since then their integrals should be subtracted because doing cycle along the compact direction $\phi$ would bring a shift in time \cite{1984JMP....25..992C},\cite{Modak_2011},\cite{PhysRevLett.39.1641}.\\
In order to avoid divergences in the Komar-like quantities, $(*dk)_{\theta t}$ and $(*dm)_{\theta t}$ have to be a good radial falloff behavior, therefore
\begin{equation}
\begin{aligned}
&(g_b)^{rr}(g_b)^{tt}(dk)_{rt} = -\frac{C}{r^2 sin^2(\theta)(g_b)_{rr}}\frac{\partial A}{\partial r} \Rightarrow \frac{C}{ sin^2(\theta)}\sqrt{\frac{C}{(g_b)_{rr}}}\frac{\partial A}{\partial r}\sim r^{n_1}f_1(\theta)\\
&(g_b)^{rr}(g_b)^{t \phi}(dk)_{r\phi} = \frac{B}{r^2 sin^2(\theta)(g_b)_{rr}}\frac{\partial C}{\partial r} \Rightarrow  \frac{B}{ sin^2(\theta)}\sqrt{\frac{C}{(g_b)_{rr}}}\frac{\partial C}{\partial r} \sim r^{n_2}f_2(\theta)\\
&(g_b)^{rr}(g_b)^{tt}(dm)_{rt} = -\frac{C}{r^2 sin^2(\theta)(g_b)_{rr}}\frac{\partial B}{\partial r} \Rightarrow \frac{C}{ sin^2(\theta)}\sqrt{\frac{C}{(g_b)_{rr}}}\frac{\partial B}{\partial r} \sim r^{n_3}f_3(\theta)\\
&(g_b)^{rr}(g_b)^{t\phi}(dm)_{r\phi} = \frac{B}{r^2 sin^2(\theta)(g_b)_{rr}}\frac{\partial C}{\partial r} \Rightarrow  \frac{B}{ sin^2(\theta)}\sqrt{\frac{C}{(g_b)_{rr}}}\frac{\partial C}{\partial r} \sim r^{n_4}f_4(\theta)\\
\end{aligned}    
\end{equation}
with 
\begin{equation}
\begin{aligned}
n_1,n_2,n_3,n_4 \leq 1 
\end{aligned}
\end{equation}
and $f_1(\theta),f_2(\theta),f_3(\theta),f_3(\theta)$ arbitrary function of azimuthal angle; we have also taken into account the square root of the induced metric determinant that enter in the computation of the Komar-like integrals $\sqrt{|\gamma|}=r\sqrt{C (g_b)_{rr}}$.

\section{Asymptotic generating matrix equations}\label{appB}
Let us start with the asymptotic expansion for poles $\lambda_k$
\begin{equation}
\begin{aligned}
    \lambda_k&=(w_k-z)\pm \sqrt{(w_k-z)^2+\rho^2}=\\
    &=(w_k-rcos(\theta))\pm \sqrt{(w_k-rcos(\theta))^2+r^2sin^2(\theta)}=\\
    &=(w_k-rcos(\theta))\pm \sqrt{(w^2_k+r^2+2w_kcos(\theta))} \sim r(-cos(\theta)\pm 1); 
\end{aligned}    
\end{equation}
Assuming the asymptotic behaviour of the spectral parameter is the same of the poles\footnote{This is suggested from the fact that the spectral parameter meets the poles $\lambda_k$ along its trajectory; if the poles are considered on the asymptotic boundary also the whole trajectory has to be consider asymptotically and it has to meet poles on the asymptotic boundary.} we have to expand the following coefficients:
\begin{equation}
\begin{aligned}
&\frac{2\lambda^2}{\lambda^2+\rho^2}\sim \frac{2(-rcos(\theta) \pm r)^2}{(-rcos(\theta)\pm r)^2+r^2sin^2(\theta)}=\frac{2[r^2cos^2(\theta)+r^2\mp 2r^2cos(\theta)]}{2r^2\mp 2r^2cos(\theta)}= \\
&=1\mp \frac{cos(\theta)}{1\mp cos(\theta)}+\frac{cos^2(\theta)}{1\mp cos(\theta)};\\
&\frac{2\lambda \rho}{\lambda^2+\rho^2}\sim \frac{2(-rcos(\theta)\pm r)rsin(\theta)}{2r^2\mp2r^2cos(\theta)}=-\frac{cos(\theta)sin(\theta)}{1\mp cos(\theta)}\pm \frac{sin(\theta)}{1\mp cos(\theta)};\\
&\frac{\rho}{\lambda^2+\rho^2}\sim \frac{rsin(\theta)}{2r^2\mp2r^2cos(\theta)}=\frac{1}{2r}\frac{sin(\theta)}{1\mp cos(\theta)};\\
&\frac{\lambda}{\lambda^2+\rho^2}\sim \frac{-rcos(\theta)\pm r}{2r^2\mp2r^2cos(\theta)}=\frac{1}{2r}\frac{-cos(\theta)\pm 1}{1\mp cos(\theta)}.
\label{B.2}
\end{aligned}  
\end{equation}
We can sum and subtract equations in system \ref{BNMb} to get 
\begin{equation}
\begin{aligned}
&\bigg([c(\theta)+s(\theta)]\frac{\partial}{\partial r}+[c(\theta)-s(\theta)]\frac{\partial}{\partial \theta}+\frac{2\lambda \rho-2\lambda^2}{\lambda^2+r^2s^2(\theta)}\partial_{\lambda}-\frac{2rs(\theta) V_b}{\lambda^2+r^2s^2(\theta)}\bigg)\psi_b=0, \\
&\bigg([c(\theta)-s(\theta)]\frac{\partial}{\partial r}-[s(\theta)+c(\theta)]\frac{\partial}{\partial \theta}-\frac{2\lambda^2 + 2\lambda \rho}{\lambda^2+r^2s^2(\theta)}\partial_{\lambda}+\frac{2\lambda U_b}{\lambda^2+r^2s^2(\theta)}\bigg)\psi_b=0;
\label{3.15}
\end{aligned}    
\end{equation}
where $c(\theta)=cos(\theta)$ and $s(\theta)=sin(\theta)$. Inserting the asymptotic expansions \ref{B.2}
\begin{equation}
\begin{aligned}
&\bigg([c(\theta)+s(\theta)]\frac{\partial}{\partial r}+[c(\theta)-s(\theta)]\frac{\partial}{\partial \theta}+\mathcal{F}^{\uparrow \downarrow}_1(\theta)\partial_{\lambda}\bigg)\psi_b=\frac{1}{r}\frac{s(\theta)}{1\mp c(\theta)}V_b\psi_b, \\
&\bigg([c(\theta)-s(\theta)]\frac{\partial}{\partial r}-[s(\theta)+c(\theta)]\frac{\partial}{\partial \theta}+\mathcal{F}^{\uparrow \downarrow}_2(\theta)\partial_{\lambda}\bigg)\psi_b=-\frac{1}{r}\frac{-c(\theta)\pm 1}{1\mp c(\theta)}U_b\psi_b;
\label{B.4}
\end{aligned}    
\end{equation}
where
\begin{equation}
\begin{aligned}
&\mathcal{F}^{\uparrow \downarrow}_1(\theta)=\frac{1\mp 2c(\theta)+c^2(\theta)+c(\theta)s(\theta)\mp s(\theta)}{1\mp c(\theta)}=\mp s(\theta) \mp c(\theta)+1;\\
&\mathcal{F}^{\uparrow \downarrow}_2(\theta)=\frac{1\mp 2c(\theta)+c^2(\theta)-c(\theta)s(\theta)\pm s(\theta)}{1\mp c(\theta)}=\pm s(\theta) \mp c(\theta)+1;
\end{aligned}    
\end{equation}
where the apexes ${\uparrow \downarrow}$ indicate respectively the up or down choice for the signs; matrices $V_b$ and $U_b$ could be expanded to get
\begin{equation}
\begin{aligned}
&V_b=\rho (g_b)_{,z}(g_b)^{-1} \sim \\
&\sim \frac{1}{rsin(\theta)}\begin{bmatrix}
r^{\gamma}\mathcal{C}sin(\theta)r^{\alpha}\mathcal{A}'-r^{\beta}\mathcal{B}r^{\beta}sin(\theta)\mathcal{B}' & \ \ \ -r^{\beta}\mathcal{B}sin(\theta)r^{\alpha}\mathcal{A}'+r^{\alpha}\mathcal{A}r^{\beta}sin(\theta)\mathcal{B}' \\
r^{\beta}\mathcal{B}'sin(\theta)r^{\gamma}\mathcal{C}-r^{\beta}\mathcal{B}r^{\gamma}sin(\theta)\mathcal{C}' & \ \ \ -r^{\beta}\mathcal{B}sin(\theta)r^{\beta}\mathcal{B}'+r^{\alpha}\mathcal{A}r^{\gamma}sin(\theta)\mathcal{C}' 
\end{bmatrix};
\end{aligned}    
\end{equation}
\begin{equation}
\begin{aligned}
&U_b=\rho (g_b)_{,\rho}(g_b)^{-1} \sim \\
& \sim \frac{1}{rsin(\theta)}\begin{bmatrix}
-r^{\gamma}\mathcal{C}cos(\theta)r^{\alpha}\mathcal{A}'+r^{\beta}\mathcal{B}r^{\beta}cos(\theta)\mathcal{B}' & \ \ \ r^{\beta}\mathcal{B}cos(\theta)r^{\alpha}\mathcal{A}'-r^{\alpha}\mathcal{A}r^{\beta}cos(\theta)\mathcal{B}' \\
-r^{\beta}\mathcal{B}'cos(\theta)r^{\gamma}\mathcal{C}+r^{\beta}\mathcal{B}r^{\gamma}cos(\theta)\mathcal{C}' & \ \ \ r^{\beta}\mathcal{B}cos(\theta)r^{\beta}\mathcal{B}'-r^{\alpha}\mathcal{A}r^{\gamma}cos(\theta)\mathcal{C}' 
\end{bmatrix}.
\end{aligned}    
\end{equation}
Note that asymptotically we have $V_b \sim cotang(\theta)U_b$.

\section{Christoffel symbols for the $N$ solitons solutions}\label{cri}
Let us derive here the non-vanishing Christoffel symbols for a metric like
\begin{equation}
    ds^2=g_{22}(dr^2+r^2d\theta^2)+g_{ij}dx^idx^j \ \ \ i,j=0,1;
\end{equation}
which is of the same kind of the solitons metric.
The non-vanishing symbols are
\begin{equation}
\begin{aligned}
    &\Gamma^2_{00}=-\frac{1}{2}g^{22}\partial_2g_{00}; \ \ \Gamma^3_{00}=-\frac{1}{2r^2}g^{22}\partial_3g_{00}; \ \ \Gamma^2_{01}=-\frac{1}{2}g^{22}\partial_2g_{01}; \ \ \Gamma^3_{01}=-\frac{1}{2r^2}g^{22}\partial_3g_{01}; \\
    & \Gamma^0_{20}=\frac{1}{2}g^{00}\partial_2g_{00}+\frac{1}{2}g^{10}\partial_2g_{01}; \ \ \Gamma^1_{20}=\frac{1}{2}g^{01}\partial_2g_{00}+\frac{1}{2}g^{11}\partial_2g_{01}; \\
    & \Gamma^0_{30}=\frac{1}{2}g^{00}\partial_3g_{00}+\frac{1}{2}g^{10}\partial_3g_{01}; \ \ \Gamma^1_{30}=\frac{1}{2}g^{01}\partial_3g_{00}+\frac{1}{2}g^{11}\partial_3g_{01}; \\
    &\Gamma^2_{11}=-\frac{1}{2}g^{22}\partial_2g_{11}; \ \ \Gamma^3_{11}=-\frac{1}{2r^2}g^{22}\partial_3g_{11}; \ \ \Gamma^0_{22}=-\frac{1}{2}g^{00}\partial_2g_{22}; \ \ \Gamma^2_{22}=-\frac{1}{2}g^{22}\partial_2g_{22};\\
     & \Gamma^0_{12}=\frac{1}{2}g^{00}\partial_2g_{01}+\frac{1}{2}g^{10}\partial_2g_{11}; \ \ \Gamma^1_{12}=\frac{1}{2}g^{01}\partial_2g_{11}+\frac{1}{2}g^{11}\partial_2g_{11}; \\
     & \Gamma^0_{13}=\frac{1}{2}g^{00}\partial_3g_{10}+\frac{1}{2}g^{10}\partial_3g_{11}; \ \ \Gamma^1_{12}=\frac{1}{2}g^{01}\partial_3g_{10}+\frac{1}{2}g^{11}\partial_3g_{11}; \\
     &\Gamma^3_{22}=-\frac{1}{2r^2}g^{22}\partial_3g_{22}; \ \ \Gamma^2_{23}=\frac{1}{2}g^{22}\partial_3g_{22};  \ \ \Gamma^3_{23}=\frac{1}{2r^2}g^{22}\partial_2g_{33};  \ \ \Gamma^2_{31}=-\frac{1}{2}g^{22}\partial_2g_{33};\\
     &\Gamma^3_{33}=\frac{1}{2r^2}g^{22}\partial_3g_{33}.
\end{aligned}
\end{equation}
The one we are more interested to deduce which is the leading order of the killing vectors is the last one and for $N=0$ we simply have
\begin{equation}
\big(\Gamma^{3}_{33}\big)_{N=0}=\frac{\mathcal{D}'(\theta)}{2\mathcal{D}(\theta)}
\end{equation}
 while for the case $N=2$ we get
\begin{equation}
\begin{aligned}
\big(\Gamma^{3}_{33}\big)_{N=2}&=\mathcal{M}_{2}(\theta)\bigg[{(r^{\epsilon_1+\epsilon_3+1}\mathcal{E}_1\mathcal{E}_3(\theta)-r^{2\epsilon_2+1}\mathcal{E}^2_2(\theta))}\bigg]^{-4} r^{\delta+2+2\#} \bigg(\prod_{j=1}^{\#}\frac{1}{(w_k-w_l)_j^2}\bigg)\Xi_2 \ \times \\
& \times \bigg[\frac{\mathcal{M}'_{2}(\theta)}{\mathcal{M}_{2}(\theta)}-4\frac{r^{\epsilon_1+\epsilon_3+1}(\mathcal{E}_1\mathcal{E}_3(\theta))'-r^{2\epsilon_2+1}(\mathcal{E}_2^2(\theta))'}{r^{\epsilon_1+\epsilon_3+1}\mathcal{E}_1\mathcal{E}_3(\theta)-r^{2\epsilon_2+1}\mathcal{E}_2^2(\theta)}+\frac{\Xi'_2}{\Xi_2}\bigg]\frac{1
}{2g^{22}}=\\
&=\frac{1}{2}\bigg[\frac{\mathcal{M}'_{2}(\theta)}{\mathcal{M}_{2}(\theta)}-4\frac{r^{\epsilon_1+\epsilon_3+1}(\mathcal{E}_1\mathcal{E}_3(\theta))'-r^{2\epsilon_2+1}(\mathcal{E}_2^2(\theta))'}{r^{\epsilon_1+\epsilon_3+1}\mathcal{E}_1\mathcal{E}_3(\theta)-r^{2\epsilon_2+1}\mathcal{E}_2^2(\theta)}+\frac{\Xi'_2}{\Xi_2}\bigg];
\label{gamman2}
\end{aligned}
\end{equation}
where $\Xi_2:=\sum _{\sigma \in S_{2}}\big(\operatorname {sgn}(\sigma)  \prod _{s=0}^{2}[{A_{s\sigma_s}}+{B_{s\sigma_s}}+{C_{s\sigma_s}}+{D_{s\sigma_s}}]\big)$. 
The first term has no power of the radial cooordinate while the second and third terms are of the form
\begin{equation}
    \frac{r^af'(\theta)+r^bg'(\theta)}{r^af(\theta)+r^bg(\theta)},
\end{equation}
where $a$ and $b$ are certain numbers; if $r^a$ is dominant with respect to $r^b$ we will get $\frac{f'(\theta)}{f(\theta)}$ and if $r^b$ is dominant with respect to $r^a$ we will get $\frac{g'(\theta)}{g(\theta)}$. Therefore, in any case, the leading order is independent of $r$ but strongly depends on the initial data\footnote{This could be interpreted as a chaotic behaviour of axialgravisolitons.}. For the general $N$ case the story is no different; we get
\begin{equation}
\begin{aligned}
\big(\Gamma^{3}_{33}\big)_{N}&=\mathcal{M}_{N}(\theta)\bigg[{(r^{\epsilon_1+\epsilon_3+1}\mathcal{E}_1\mathcal{E}_3(\theta)-r^{2\epsilon_2+1}\mathcal{E}^2_2(\theta))}\bigg]^{-2N} r^{\bar{\delta}} \bigg(\prod_{j=1}^{\#}\frac{1}{(w_k-w_l)_j^2}\bigg)\Xi_N \ \times \\
& \times \bigg[\frac{\mathcal{M}'_{N}(\theta)}{\mathcal{M}_{N}(\theta)}-2N\frac{r^{\epsilon_1+\epsilon_3+1}(\mathcal{E}_1\mathcal{E}_3(\theta))'-r^{2\epsilon_2+1}(\mathcal{E}_2^2(\theta))'}{r^{\epsilon_1+\epsilon_3+1}\mathcal{E}_1\mathcal{E}_3(\theta)-r^{2\epsilon_2+1}\mathcal{E}_2^2(\theta)}+\frac{\Xi'_N}{\Xi_N}\bigg]\frac{1
}{2g^{22}}=\\
&=\frac{1}{2}\bigg[\frac{\mathcal{M}'_{N}(\theta)}{\mathcal{M}_{N}(\theta)}-2N\frac{r^{\epsilon_1+\epsilon_3+1}(\mathcal{E}_1\mathcal{E}_3(\theta))'-r^{2\epsilon_2+1}(\mathcal{E}_2^2(\theta))'}{r^{\epsilon_1+\epsilon_3+1}\mathcal{E}_1\mathcal{E}_3(\theta)-r^{2\epsilon_2+1}\mathcal{E}_2^2(\theta)}+\frac{\Xi'_N}{\Xi_N}\bigg];
\label{gammann}
\end{aligned}
\end{equation}
where now $\Xi_N:=\sum _{\sigma \in S_{N}}\big(\operatorname {sgn}(\sigma)  \prod _{s=0}^{N}[{A_{s\sigma_s}}+{B_{s\sigma_s}}+{C_{s\sigma_s}}+{D_{s\sigma_s}}]\big)$, $\bar{\delta}:=\delta-\frac{N^2}{2}+N(N+1)-N(N-1)+2\#$ and the same consideration above for the case $N=2$ can be applied.

\bibliographystyle{elsarticle-num} 
\bibliography{biblio}

\begin{thebibliography}{10}
\expandafter\ifx\csname url\endcsname\relax
  \def\url#1{\texttt{#1}}\fi
\expandafter\ifx\csname urlprefix\endcsname\relax\def\urlprefix{URL }\fi
\expandafter\ifx\csname href\endcsname\relax
  \def\href#1#2{#2} \def\path#1{#1}\fi

\bibitem{grenne}
C.~S. Gardner, J.~M. Greene, M.~D. Kruskal, R.~M. Miura,
  \href{https://link.aps.org/doi/10.1103/PhysRevLett.19.1095}{{Method for
  Solving the Korteweg-deVries Equation}}, Phys. Rev. Lett. 19 (1967)
  1095--1097.
\newblock \href {https://doi.org/10.1103/PhysRevLett.19.1095}
  {\path{doi:10.1103/PhysRevLett.19.1095}}.
\newline\urlprefix\url{https://link.aps.org/doi/10.1103/PhysRevLett.19.1095}

\bibitem{shabat}
V.~E. {Zakharov}, A.~B. {Shabat}, {Exact Theory of Two-dimensional
  Self-focusing and One-dimensional Self-modulation of Waves in Nonlinear
  Media}, Soviet Journal of Experimental and Theoretical Physics 34 (1972) 62.

\bibitem{1978ZhETF..75.1953B}
V.~A. {Belinskii}, V.~E. {Zakharov}, {Integration of the Einstein equations by
  the method of the inverse scattering problem and calculation of exact soliton
  solutions}, Zhurnal Eksperimentalnoi i Teoreticheskoi Fiziki 75 (1978)
  1953--1971.

\bibitem{1980JETPL..32..277A}
G.~A. {Alekseev}, {N-soliton solutions of Einstein-Maxwell equations}, Soviet
  Journal of Experimental and Theoretical Physics Letters 32 (1980) 277--279.

\bibitem{Vigano:2022hrg}
A.~Vigan\`o, {Black Holes and Solution Generating Techniques}, Ph.D. thesis,
  Milan U. (2022).
\newblock \href {http://arxiv.org/abs/2211.00436} {\path{arXiv:2211.00436}}.

\bibitem{Asselmeyer_Maluga_2020}
T.~Asselmeyer-Maluga, J.~Król,
  \href{http://dx.doi.org/10.3390/universe6120234}{Dark matter as gravitational
  solitons in the weak field limit}, Universe 6~(12) (2020) 234.
\newblock \href {https://doi.org/10.3390/universe6120234}
  {\path{doi:10.3390/universe6120234}}.
\newline\urlprefix\url{http://dx.doi.org/10.3390/universe6120234}

\bibitem{mcnamara2021gravitational}
J.~McNamara, Gravitational solitons and completeness (2021).
\newblock \href {http://arxiv.org/abs/2108.02228} {\path{arXiv:2108.02228}}.

\bibitem{kordas2011transition}
P.~Kordas, Transition matrix, poisson bracket for gravitational solitons in the
  dressing formalism (2011).
\newblock \href {http://arxiv.org/abs/1002.0524} {\path{arXiv:1002.0524}}.

\bibitem{durgut2022supersymmetric}
T.~Durgut, H.~K. Kunduri, Supersymmetric multi-charge solitons in ads$_5$
  (2022).
\newblock \href {http://arxiv.org/abs/2111.06831} {\path{arXiv:2111.06831}}.

\bibitem{Durgut_2023}
T.~Durgut, H.~K. Kunduri,
  \href{http://dx.doi.org/10.1016/j.aop.2023.169435}{Supersymmetric
  asymptotically locally ads<mml:math
  xmlns:mml="http://www.w3.org/1998/math/mathml" display="inline" id="d1e22"
  altimg="si9.svg"><mml:msub><mml:mrow
  /><mml:mrow><mml:mn>5</mml:mn></mml:mrow></mml:msub></mml:math> gravitational
  solitons}, Annals of Physics 457 (2023) 169435.
\newblock \href {https://doi.org/10.1016/j.aop.2023.169435}
  {\path{doi:10.1016/j.aop.2023.169435}}.
\newline\urlprefix\url{http://dx.doi.org/10.1016/j.aop.2023.169435}

\bibitem{Durgut_20232}
T.~Durgut, R.~A. Hennigar, H.~K. Kunduri, R.~B. Mann,
  \href{http://dx.doi.org/10.1007/JHEP03(2023)114}{Phase transitions and
  stability of eguchi-hanson-ads solitons}, Journal of High Energy Physics
  2023~(3) (Mar. 2023).
\newblock \href {https://doi.org/10.1007/jhep03(2023)114}
  {\path{doi:10.1007/jhep03(2023)114}}.
\newline\urlprefix\url{http://dx.doi.org/10.1007/JHEP03(2023)114}

\bibitem{Bondi:1962px}
H.~Bondi, M.~G.~J. van~der Burg, A.~W.~K. Metzner, {Gravitational waves in
  general relativity. 7. Waves from axisymmetric isolated systems}, Proc. Roy.
  Soc. Lond. A 269 (1962) 21--52.
\newblock \href {https://doi.org/10.1098/rspa.1962.0161}
  {\path{doi:10.1098/rspa.1962.0161}}.

\bibitem{Sachs:1962wk}
R.~K. Sachs, {Gravitational waves in general relativity. 8. Waves in
  asymptotically flat space-times}, Proc. Roy. Soc. Lond. A 270 (1962)
  103--126.
\newblock \href {https://doi.org/10.1098/rspa.1962.0206}
  {\path{doi:10.1098/rspa.1962.0206}}.

\bibitem{PhysRev.128.2851}
R.~Sachs, \href{https://link.aps.org/doi/10.1103/PhysRev.128.2851}{Asymptotic
  symmetries in gravitational theory}, Phys. Rev. 128 (1962) 2851--2864.
\newblock \href {https://doi.org/10.1103/PhysRev.128.2851}
  {\path{doi:10.1103/PhysRev.128.2851}}.
\newline\urlprefix\url{https://link.aps.org/doi/10.1103/PhysRev.128.2851}

\bibitem{Compere:2018aar}
G.~Comp\`ere, A.~Fiorucci, {Advanced Lectures on General Relativity} (1 2018).
\newblock \href {http://arxiv.org/abs/1801.07064} {\path{arXiv:1801.07064}}.

\bibitem{Ferrero:2024eva}
P.~Ferrero, D.~Francia, C.~Heissenberg, M.~Romoli, {Double-Copy
  Supertranslations} (2 2024).
\newblock \href {http://arxiv.org/abs/2402.11595} {\path{arXiv:2402.11595}}.

\bibitem{Strominger:2017aeh}
A.~Strominger, {Black Hole Information Revisited}, 2020.
\newblock \href {http://arxiv.org/abs/1706.07143} {\path{arXiv:1706.07143}},
  \href {https://doi.org/10.1142/9789811203961_0010}
  {\path{doi:10.1142/9789811203961_0010}}.

\bibitem{Francia:2018jtb}
D.~Francia, C.~Heissenberg, {Two-Form Asymptotic Symmetries and Scalar Soft
  Theorems}, Phys. Rev. D 98~(10) (2018) 105003.
\newblock \href {http://arxiv.org/abs/1810.05634} {\path{arXiv:1810.05634}},
  \href {https://doi.org/10.1103/PhysRevD.98.105003}
  {\path{doi:10.1103/PhysRevD.98.105003}}.

\bibitem{Campoleoni:2018uib}
A.~Campoleoni, D.~Francia, C.~Heissenberg, {Asymptotic symmetries and charges
  at null infinity: from low to high spins}, EPJ Web Conf. 191 (2018) 06011.
\newblock \href {http://arxiv.org/abs/1808.01542} {\path{arXiv:1808.01542}},
  \href {https://doi.org/10.1051/epjconf/201819106011}
  {\path{doi:10.1051/epjconf/201819106011}}.

\bibitem{Pasterski:2015zua}
S.~Pasterski, {Asymptotic Symmetries and Electromagnetic Memory}, JHEP 09
  (2017) 154.
\newblock \href {http://arxiv.org/abs/1505.00716} {\path{arXiv:1505.00716}},
  \href {https://doi.org/10.1007/JHEP09(2017)154}
  {\path{doi:10.1007/JHEP09(2017)154}}.

\bibitem{Henneaux:1986ht}
M.~Henneaux, C.~Teitelboim, {P FORM ELECTRODYNAMICS}, Found. Phys. 16 (1986)
  593--617.
\newblock \href {https://doi.org/10.1007/BF01889624}
  {\path{doi:10.1007/BF01889624}}.

\bibitem{Afshar:2018apx}
H.~Afshar, E.~Esmaeili, M.~M. Sheikh-Jabbari, {Asymptotic Symmetries in
  $p$-Form Theories}, JHEP 05 (2018) 042.
\newblock \href {http://arxiv.org/abs/1801.07752} {\path{arXiv:1801.07752}},
  \href {https://doi.org/10.1007/JHEP05(2018)042}
  {\path{doi:10.1007/JHEP05(2018)042}}.

\bibitem{Esmaeili:2020eua}
E.~Esmaeili, {$p$-form gauge fields: charges and memories}, Ph.D. thesis, IPM,
  Tehran (9 2020).
\newblock \href {http://arxiv.org/abs/2010.13922} {\path{arXiv:2010.13922}}.

\bibitem{manz1}
Francia, Heissenberg, Manzoni, {Asymptotic charges of p-form and their duals.
  To be announced}.

\bibitem{manz2}
F.~Manzoni, {A mathematical note on the asymptotic symmetries of p-forms and
  their duals. To be announced}.

\bibitem{Speranza:2017gxd}
A.~J. Speranza, {Local phase space and edge modes for diffeomorphism-invariant
  theories}, JHEP 02 (2018) 021.
\newblock \href {http://arxiv.org/abs/1706.05061} {\path{arXiv:1706.05061}},
  \href {https://doi.org/10.1007/JHEP02(2018)021}
  {\path{doi:10.1007/JHEP02(2018)021}}.

\bibitem{Geiller:2017whh}
M.~Geiller, {Lorentz-diffeomorphism edge modes in 3d gravity}, JHEP 02 (2018)
  029.
\newblock \href {http://arxiv.org/abs/1712.05269} {\path{arXiv:1712.05269}},
  \href {https://doi.org/10.1007/JHEP02(2018)029}
  {\path{doi:10.1007/JHEP02(2018)029}}.

\bibitem{Donnelly:2020xgu}
W.~Donnelly, L.~Freidel, S.~F. Moosavian, A.~J. Speranza, {Gravitational edge
  modes, coadjoint orbits, and hydrodynamics}, JHEP 09 (2021) 008.
\newblock \href {http://arxiv.org/abs/2012.10367} {\path{arXiv:2012.10367}},
  \href {https://doi.org/10.1007/JHEP09(2021)008}
  {\path{doi:10.1007/JHEP09(2021)008}}.

\bibitem{Ciambelli:2021vnn}
L.~Ciambelli, R.~G. Leigh, {Isolated surfaces and symmetries of gravity}, Phys.
  Rev. D 104~(4) (2021) 046005.
\newblock \href {http://arxiv.org/abs/2104.07643} {\path{arXiv:2104.07643}},
  \href {https://doi.org/10.1103/PhysRevD.104.046005}
  {\path{doi:10.1103/PhysRevD.104.046005}}.

\bibitem{Ciambelli:2021nmv}
L.~Ciambelli, R.~G. Leigh, P.-C. Pai, {Embeddings and Integrable Charges for
  Extended Corner Symmetry}, Phys. Rev. Lett. 128 (2022).
\newblock \href {http://arxiv.org/abs/2111.13181} {\path{arXiv:2111.13181}},
  \href {https://doi.org/10.1103/PhysRevLett.128.171302}
  {\path{doi:10.1103/PhysRevLett.128.171302}}.

\bibitem{Canepa:2022uii}
G.~Canepa, A.~S. Cattaneo, {Corner Structure of Four-Dimensional General
  Relativity in the Coframe Formalism} (2 2022).
\newblock \href {http://arxiv.org/abs/2202.08684} {\path{arXiv:2202.08684}}.

\bibitem{ciambelli2023asymptotic}
L.~Ciambelli, From asymptotic symmetries to the corner proposal (2023).
\newblock \href {http://arxiv.org/abs/2212.13644} {\path{arXiv:2212.13644}}.

\bibitem{Belinski:2001ph}
V.~Belinski, E.~Verdaguer, {Gravitational solitons}, Cambridge Monographs on
  Mathematical Physics, Cambridge University Press, 2005.
\newblock \href {https://doi.org/10.1017/CBO9780511535253}
  {\path{doi:10.1017/CBO9780511535253}}.

\bibitem{Manzoni:2021dij}
F.~Manzoni, {Solitonic solutions and gravitational solitons: an overview} (2
  2021).
\newblock \href {http://arxiv.org/abs/2102.11259} {\path{arXiv:2102.11259}}.

\bibitem{Bakas:1996gs}
I.~Bakas, {2-D gravisolitons in string theory}, in: {2nd International Sakharov
  Conference on Physics}, 1996, pp. 350--354.
\newblock \href {http://arxiv.org/abs/hep-th/9606030}
  {\path{arXiv:hep-th/9606030}}.

\bibitem{strominger2018lectures}
A.~Strominger, Lectures on the infrared structure of gravity and gauge theory
  (2018).
\newblock \href {http://arxiv.org/abs/1703.05448} {\path{arXiv:1703.05448}}.

\bibitem{Mandal:2019bdu}
S.~Mandal, S.~Banerjee, {Local description of S-matrix in quantum field theory
  in curved spacetime using Riemann-normal coordinate}, Eur. Phys. J. Plus
  136~(10) (2021) 1064.
\newblock \href {http://arxiv.org/abs/1908.06717} {\path{arXiv:1908.06717}},
  \href {https://doi.org/10.1140/epjp/s13360-021-02037-z}
  {\path{doi:10.1140/epjp/s13360-021-02037-z}}.

\bibitem{Pasterski_2016}
S.~Pasterski, A.~Strominger, A.~Zhiboedov,
  \href{http://dx.doi.org/10.1007/JHEP12(2016)053}{New gravitational memories},
  Journal of High Energy Physics 2016~(12) (Dec. 2016).
\newblock \href {https://doi.org/10.1007/jhep12(2016)053}
  {\path{doi:10.1007/jhep12(2016)053}}.
\newline\urlprefix\url{http://dx.doi.org/10.1007/JHEP12(2016)053}

\bibitem{Strominger:2014pwa}
A.~Strominger, A.~Zhiboedov, {Gravitational Memory, BMS Supertranslations and
  Soft Theorems}, JHEP 01 (2016) 086.
\newblock \href {http://arxiv.org/abs/1411.5745} {\path{arXiv:1411.5745}},
  \href {https://doi.org/10.1007/JHEP01(2016)086}
  {\path{doi:10.1007/JHEP01(2016)086}}.

\bibitem{1984JMP....25..992C}
J.~M. {Cohen}, F.~{de Felice}, {The total effective mass of the Kerr-Newman
  metric}, Journal of Mathematical Physics 25~(4) (1984) 992--994.
\newblock \href {https://doi.org/10.1063/1.526217}
  {\path{doi:10.1063/1.526217}}.

\bibitem{Modak_2011}
S.~K. Modak, S.~Samanta,
  \href{https://doi.org/10.1007%2Fs10773-011-1017-2}{Effective values of komar
  conserved quantities and their applications}, International Journal of
  Theoretical Physics 51~(5) (2011) 1416--1424.
\newblock \href {https://doi.org/10.1007/s10773-011-1017-2}
  {\path{doi:10.1007/s10773-011-1017-2}}.
\newline\urlprefix\url{https://doi.org/10.1007%2Fs10773-011-1017-2}

\bibitem{PhysRevLett.39.1641}
J.~M. Cohen, H.~E. Moses,
  \href{https://link.aps.org/doi/10.1103/PhysRevLett.39.1641}{New test of the
  synchronization procedure in noninertial systems}, Phys. Rev. Lett. 39 (1977)
  1641--1643.
\newblock \href {https://doi.org/10.1103/PhysRevLett.39.1641}
  {\path{doi:10.1103/PhysRevLett.39.1641}}.
\newline\urlprefix\url{https://link.aps.org/doi/10.1103/PhysRevLett.39.1641}

\end{thebibliography}

\end{document}